\documentclass[journal,10pt]{IEEEtran}
\newif\ifCLASSOPTIONromanappendices \CLASSOPTIONromanappendicestrue

\usepackage{a0size}
\usepackage{amssymb}
\usepackage{multicol}
\usepackage[english]{babel}
\usepackage{epsfig}
\usepackage{bm}
\usepackage{amsfonts,color,amsthm,amsmath, lscape,graphics}
\usepackage{url}
\usepackage{algpseudocode}

\usepackage[font=footnotesize]{caption}
\usepackage{subcaption}
\usepackage{epstopdf}
\usepackage[ruled]{algorithm2e}
\usepackage{algpseudocode}
\usepackage{cite}
\usepackage{relsize}
\usepackage{fix2col}
\usepackage{psfrag}
\DeclareMathOperator*{\Maximize}{maximize}

\renewcommand{\figurename}{Fig.}
\addto\captionsenglish{\renewcommand{\figurename}{Fig.}}

\newcommand{\bh}{\mathbf{h}}

\newcommand{\bv}{\mathbf{v}}
\newcommand{\bV}{\mathbf{V}}

\newcommand{\bI}{\mathbf{I}}

\newcommand{\br}{\mathbf{r}}

\newcommand{\bH}{\mathbf{H}}

\renewcommand{\frac}{\dfrac}

\newcommand{\norm}[1]{\left\lVert#1\right\rVert}

\DeclareMathOperator*{\minimize}{minimize}
\DeclareMathOperator*{\diag}{diag}

\newcommand{\Tr}{{\mbox{Tr}}}

\definecolor{myOrange}{rgb}{1,0.5,0}
\definecolor{myGreen}{rgb}{0,0.5,0}
\newcommand{\editb}[1]{{\color{black}#1}}
\newcommand{\editbb}[1]{{\color{black}#1}}
\newcommand{\editbbb}[1]{{\color{black}#1}}
\newcommand{\editbl}[1]{{\color{black}#1}}
\newcommand{\editll}[1]{{\color{black}#1}}
\newcommand{\editr}[1]{{\color{black}#1}}
\newcommand{\editrr}[1]{{\color{black}#1}}
\newcommand{\editt}[1]{{\color{black}#1}}
\newcommand{\editf}[1]{{\color{black}#1}}
\newcommand{\editff}[1]{{\color{black}#1}}
\newcommand{\edit}[1]{{\color{black}#1}}
\newcommand{\editg}[1]{{\color{black}#1}}
\newcommand{\editgg}[1]{{\color{black}#1}}
\newcommand{\editggg}[1]{{\color{black}#1}}
\newcommand{\edittt}[1]{{\color{black}#1}}
\newcommand{\editrev}[1]{{\color{black}#1}}

\newcommand{\editrevrr}[1]{{\color{black}#1}}
\newcommand{\editfi}[1]{{\color{black}#1}}

\begin{document}
%
\title{Deep Learning for Channel Sensing and Hybrid Precoding in TDD Massive MIMO \editrr{OFDM Systems}}
%
%
%
\author{Kareem~M.~Attiah,~\IEEEmembership{Graduate~Student~Member,~IEEE,} Foad~Sohrabi,~\IEEEmembership{Member,~IEEE,} and~Wei~Yu,~\IEEEmembership{Fellow,~IEEE}
\thanks{
The authors are with The Edward S.\ Rogers Sr.\ Department of Electrical and Computer Engineering, University of Toronto, Toronto, ON M5S 3G4, Canada (e-mails:\{kattiah, fsohrabi, weiyu\}@ece.utoronto.ca). The materials in this paper have been presented in part at IEEE Global Communications Conference (Globecom) Workshop, Taipei, Taiwan, December 2020 \cite{attiahdeep2020}. This work was supported by 
Huawei Technologies Canada. 
}
}
\maketitle

\begin{abstract}
This paper proposes a deep learning approach to channel sensing and downlink hybrid
beamforming for massive multiple-input multiple-output systems operating in the time division
duplex mode and employing either single-carrier or multicarrier transmission. The
conventional precoding design involves a two-step process of first estimating the high-dimensional channel, then designing the precoders based on such estimate. This two-step
process is, however, not necessarily optimal. This paper shows that by using a learning
approach to design the analog sensing and the hybrid downlink precoders directly from the
received pilots without the intermediate high-dimensional channel estimation, the overall
system performance can be significantly improved. Training a neural network to design the analog
and digital precoders simultaneously is, however, difficult. Further, such an approach is not generalizable to systems with different number of users. In this paper, we develop a simplified and generalizable approach that learns the uplink sensing
matrix and downlink analog precoder using a deep neural network that decomposes on a per-user basis, then designs the digital precoder
based on the estimated low-dimensional equivalent channel. Numerical comparisons show
that the proposed methodology results in significantly less training overhead and leads to an
architecture that generalizes to various system settings.
\end{abstract}

\begin{IEEEkeywords}
Deep learning, hybrid precoding, massive multiple-input multiple-output (MIMO), millimeter wave (mmWave), time division duplex (TDD).
\end{IEEEkeywords}

\section{Introduction}
Millimeter wave (mmWave) communication 
\editll{has} attracted significant interest as means of addressing the increasing demand for faster data rates in future cellular networks\cite{rangan2014millimeter, heathoverview, wangmulti2014}. \editll{As compared to} the traditional sub-$30$GHz communication bands, wireless transmissions at mmWave frequencies experience 
\editll{more severe path and penetration loss.} Fortunately, \editll{the poor channel conditions of mmWave communications} 
can be effectively mitigated through the use of massive multiple-input multiple-output (MIMO) antenna arrays~\cite{Marzetta2010}. 
\editt{However,} despite the advantages of utilizing large antenna arrays, the practical deployment of the fully digital massive MIMO \editr{is \editt{hindered} by the excessive power consumption associated with the \editt{large} number of radio frequency (RF) chains.} 
To overcome this limitation, several alternative solutions have been proposed 
to permit the use of large antenna arrays \editll{while reducing} 
the high power \editll{consumption.} 
In this paper, we focus on the so-called hybrid beamforming architecture~\cite{zhang2005variable,el2014spatially} wherein the fully digital beamformer is replaced by \editll{an} analog beamformer that maps the received signal on the antennas into a small number of RF chains using a network of phase shifters, followed by a low-dimensional digital beamformer.

This paper \editll{considers} 
the problem of constructing the hybrid beamformers from imperfect channel state information (CSI) in massive MIMO systems for frequency-flat \editll{as well as} frequency-selective propagation environments with \editll{a} limited number of scatters. We focus on \edit{the time-division duplex (TDD)} operation so that channel reciprocity can be exploited in order to efficiently acquire the CSI at the base station (BS) using uplink pilots~\cite{Marzetta2010}. 
The existing hybrid precoding strategies for TDD massive MIMO systems follow a two-step methodology that decomposes the overall precoding procedure into a channel sensing and estimation step \editll{followed by} a subsequent downlink precoding step.   
In the channel sensing and estimation step, the spatial and frequency characteristics of the mmWave channel are often exploited to facilitate the estimation of the channel parameters, e.g.,~\cite{Lee2016,alkhateeb2014channel,Bellili2019GAMP, Rodriguez2018}.
\editr{Then}, in the downlink precoding step, the BS \editr{constructs} the precoding matrices using 
algorithms that \editll{treat the estimated CSI as} perfect CSI. 

This work 
is motivated by the key observation 
that the \editt{above} conventional paradigm is not necessarily optimal, especially in the short pilot regime. \editt{This is because} 
the \edit{conventional} channel estimation 
\edit{process typically uses a specific distance metric, such as the square loss \editg{(possibly with a regularizer on the model parameters),} 
without accounting} for the effect of channel estimation errors in the subsequent precoding step. Consequently, such a metric may not exactly match the ultimate goal of maximizing the overall system performance (e.g., sum rate). The main \editll{point} of this work is that by adopting an end-to-end design that directly \editll{constructs} the hybrid precoders from the received pilots without the intermediate channel estimation step
, \editll{we} can overcome the drawbacks of the conventional precoding framework, thereby \editf{enhancing} the overall system performance. 

\subsection{Main Contribution\editll{s}}
This paper proposes a joint channel sensing and downlink precoding \editll{approach} that bypasses explicit channel estimation. 
Driven by the success of deep learning in tackling intricate optimization problems, we advocate the use of \editll{a} deep neural network (DNN) to model the end-to-end massive MIMO system encompassing channel sensing, estimation\editll{,} and downlink precoding\editll{,} and to learn a direct mapping for the hybrid precoding matrices. \editt{\editf{A} main contribution of this paper is that we highlight the inherent limitation of the conventional channel estimation based precoding and demonstrate how a learning based strategy is well suited to bypass this limitation.}


The first part of this work considers 
a frequency-flat mmWave propagation model. In this case, the precoding problem involves the design of the analog precoding matrix and a single digital precoding matrix. \editll{\editr{However, it turns out that} using the naive approach that simultaneously learns the analog and digital precoders is infeasible due to high training complexity and the fact that it is not easily generalizable to systems with different number of users.} 
To circumvent these limitations, 
we develop an alternative semi-direct strategy that bypasses explicit channel estimation for the analog precoding design. \editrev{In particular, we decompose 
the overall precoding design into two phases in which we separately construct the 
analog and digital precoding matrices. In the analog precoding design phase, 
a DNN is employed to map the received pilots directly into the analog precoders.} 
\editrr{The proposed DNN architecture decomposes across the users and the per-user DNNs have weights tied together in order to simplify the training complexity. 
Furthermore, to avoid the metric mismatch problem, the training of this DNN is performed so as to minimize a simplified loss function derived from the rate expression.} 
\editll{Once} the analog beamforming matrix is determined, the end-to-end hybrid precoding
system can be transformed into a low-dimensional fully digital system, thus
allowing the digital precoding matrix to be designed based on \editr{the estimated equivalent channel from a few additional pilots} using conventional techniques. \editll{Numerical results} indicate that 
\editll{the proposed} approach 
significantly outperforms the conventional framework that separates channel estimation and downlink precoding.

The second part of this paper is concerned with 
a frequency-selective propagation environment. In this case, the BS utilizes multicarrier techniques, e.g., orthogonal frequency division multiplexing (OFDM), to transmit several data streams over different subcarriers. A major challenge in the design of hybrid precoders \editll{for OFDM systems} is that the analog precoder is common to the channels across  subcarriers, whereas the digital precoder is different for every subcarrier. This physical limitation suggests a different treatment for the analog and digital design stages and can be seen as an additional motivation behind the proposed scheme of separating the designs of the analog and digital precoders. 
Specifically, 
\editll{such decoupling} allows the previous semi-direct approach to be extended naturally, i.e., we design the analog precoder using a DNN based on the \editll{received pilots} on all subcarriers \editll{and} design the digital precoders for every subcarrier based on the corresponding \editll{estimated} equivalent channel. \editll{Further, to ensure modest training complexity that does not grow with the number of subcarriers, we incorporate a convolutional stage in the neural network architecture that offers dimensionality reduction and summarizes the information in correlated received pilots over different subcarriers.}


In practice, a wireless environment is inherently non-static and the channel parameters are always changing. A major concern in \editll{using} data-driven solutions in communication systems is \editll{how robust such solutions are to changes} 
in \editll{system parameters}. 
As such, this paper also investigates the generalizability of the proposed approach in 
\editll{different} system parameters. We use experimental case studies \editll{with varying number of channel paths, uplink signal-to-noise ratio (SNR), and number of users,} 
to demonstrate the ability of the proposed scheme to maintain good performance even \editt{when the training and test sets are different}. \editll{Further, we also provide a \editt{realistic} single-cell simulation \editt{and use} a general utility function (e.g., weighted sum rate) as a means to measure the \editt{system-level} performance. The numerical findings indicate that the proposed approach is \editt{able to provide fairness among the different users.}} 

\subsection{\editll{Related Works}}
Earlier works on hybrid precoding design \editg{typically} 
assume full CSI knowledge at the BS. 
The main focus of these works is to devise low-complexity algorithms that can approach the performance of the fully digital massive MIMO system \editll{\editt{for both cases of} single-carrier \editt{and} OFDM transmissions}. \editll{\editt{Most existing} algorithms involve heuristic designs for the analog precoder and conventional linear precoding for the digital precoder. Examples of these analog heuristic designs include}  
matching to the channel phases~\cite{Li2017SVD, liang2014low}, matching to the channel strongest path~\cite{liang2014approach}, \editll{and} iterative coordinate ascent~\cite{sohrabi2016hybrid} for single-carrier systems \editll{as well as} channel covariance averaging~\cite{SohrabiOFDM, parkdynamic2017} \editll{and} time-domain matched filtering~\cite{Payami2019low} for multicarrier systems. Another interesting line of work is 
presented in~\cite{yu2016alternating} \editr{in which the precoder for a fully digital system is decomposed into analog and digital precoders using alternating minimization.} 

In practice, the CSI is not readily available at the BS and it must be estimated. The most widely common approach \editll{for channel estimation} is to take advantage of the sparsity of the mmWave \editr{channels} in the angular domain~\cite{Lee2016,alkhateeb2014channel,Bellili2019GAMP,Rodriguez2018, gao2016channel}. For the single-carrier setup,~\cite{Lee2016} develops a channel estimation algorithm inspired by the greedy orthogonal matching pursuit (OMP), while~\cite{alkhateeb2014channel} presents a multi-resolution codebook design for channel sensing and parameter estimation. Further, the work in~\cite{Bellili2019GAMP} formulates the channel estimation as a sparse recovery problem from noisy measurements, for which the generalized approximate message passing (GAMP) is used to retrieve the channel parameters.  For the multicarrier setup,~\cite{gao2016channel} develops a matching pursuit algorithm for environments \editll{with line-of-sight conditions}, whereas~\cite{Rodriguez2018} further exploits the frequency correlation of the channel across subcarriers to devise a low-complexity variant of the OMP, named simultaneous weighted OMP (SW-OMP).

Recently, \editt{deep learning} has attracted significant interest in \editll{the} wireless research community. In most hybrid precoding works, the domain of application for \editt{deep learning} is restricted to replacing key system components with a neural network. This includes replacing channel estimation~\cite{sohrabi2021beam, borgerding2017,MALAMP2021}, downlink precoding with the input being either perfect/imperfect CSI~\editrevrr{\cite{lin2020,Elbir2020, AS}}, or both~\cite{ma2020}. 
The main limitation of all these works is that they admit the traditional separation \editf{of channel estimation and precoding} and do not exploit the ability of data-driven approaches to model the end-to-end system. \editrev{For instance, in~\cite{ma2020} two separate neural networks are employed for channel estimation and analog downlink precoding. Therefore, it still adopts the traditional approach that separates channel estimation and precoding. In contrast, our work proposes to combine both modules into one that maps the received pilots directly to the analog precoding matrix.} 
\editll{We remark that} several \editt{recent} works have proposed the idea of bypassing channel estimation \editll{in \edit{related} wireless communication settings} in order to optimize some system-wide objective, e.g., \edit{for} 
scheduling in {ad-hoc} networks~\cite{cui2019spatial}, \edit{and for transmission design in systems involving} intelligent reflective surfaces~\cite{jiang2021learning} \edit{and} fully digital massive MIMO in frequency division duplex \edit{mode}~\cite{sohrabi2021deep}.

\editrevrr{Finally, we remark that the prior works~\cite{elbir2022, sun2019beam} also propose to design the hybrid precoders from the received pilots. However, the main focus of these works is quite distinct from ours. In~\cite{elbir2022}, only a single-user setup is considered, whereas the work presented herein considers the more challenging multiuser setup where the hybrid precoders are designed in the presence of user interference.  Moreover, the main goal of~\cite{elbir2022} is to mimic the solution of the existing precoding
design using supervised training while alleviating the computational burden after training. In
contrast, the proposed work aims at end-to-end modelling of the mmWave system using a realistic
system-level objective such as the overall achievable rate.
Likewise,~\cite{sun2019beam} considers a beam selection method to design the analog precoder from a fixed codebook of finite size using a non-trainable approach, whereas the present work investigates the use of deep learning to construct a direct mapping for the analog precoder without restricting
to a certain codebook construction. In addition to these works,} the very recent and parallel work~\cite{gao2022data} considers the multiuser setup, but proposes a learning-based design with a training complexity that is only suitable when the number of users is fixed and small. In particular,  the dimensions of some layers in the neural network architecture of~\cite{gao2022data} grow quadratically in the number of users. 
In contrast, the proposed solution maintains the same layer dimensions and training complexity regardless of the number of users, and can further generalize to an arbitrary number of such users without retraining. 
\IEEEpeerreviewmaketitle
\subsection{Paper Organization and Notation}
The rest of the paper is organized as follows. Section~\ref{sec:sys_mdl} presents the massive MIMO system model \editll{with hybrid architecture} in its most general form as an OFDM system assuming a frequency-selective propagation environment. 
Section III considers the special case of a single-carrier setup and develops the proposed semi-direct approach that bypasses explicit channel estimation for the analog precoding design. We then discuss the extension of the proposed approach to the multicarrier setup in Section~\ref{sec:OFDM}. In Section~\ref{sec:numerical}, we provide extensive numerical simulations that demonstrate the performance of the proposed precoding approach and examine its generalizability aspects. Finally, \editll{conclusions are drawn} in Section~\ref{sec:conclusion}.

\editr{This paper uses} lower case letters, lower case boldface letters, and upper case boldface letters to denote scalars, vectors, and matrices\editll{,} respectively. We use $[\cdot]_i$, $[\cdot]_{ij}$ to denote the $i$-th element of a vector and the element in the $i$-th row and $j$-th column of a matrix. We use $\diag{\left(\mathbf{A}_1, \ldots, \mathbf{A}_n\right)}$ to denote a block diagonal matrix with the matrices $\mathbf{A}_1, \ldots, \mathbf{A}_n$ on the diagonal. Further, $\mathbb{C}^{m, n}$ denotes an $m$
by $n$ dimensional complex space, and $\mathcal{CN}(\mathbf{0}, \mathbf{R})$ represents the
zero-mean circularly symmetric complex Gaussian distribution
with covariance matrix $\mathbf{R}$,
In addition, $\otimes$ is the Kronecker product operator,  and $\left(\cdot\right)^H$ denote\editll{s} the Hermitian transpose of matrices. The operators $|\cdot|$, $\|\cdot\|$, $\log_2{(\cdot)}$, $\mathbb{E}\left[\cdot\right],$ and $\Tr{(\cdot)}$ represent 
the absolute \editll{value}, $\ell_2$ norm, base-$2$ logarithm, expectation\editll{,} and trace, respectively. Finally, $\mathbf{I}_M$ is the identity matrix \editll{of size $M$}.

\section{Preliminaries and System Model}
\label{sec:sys_mdl}
\subsection{System Model}
Consider a TDD
massive MIMO system operating in a frequency-selective 
mmWave environment in which a BS with $M$ antennas and \editb{$N_\text{RF} < M$} RF chains \editb{\editbl{employs} OFDM \editll{transmission} over $N_c$ subcarriers to serve $K$ single-antenna users}. \editbb{Since the number of RF chains at the BS is limited}, downlink precoding is split between the analog and digital domains. Specifically, let $s_k[j]$ be the intended symbol for user $k$ over subcarrier $j$. To send a downlink data stream of symbols $\mathbf{s}[j] = \left[s_1[j], \ldots, s_K[j] \right]^T$, 
the BS precodes the \editb{symbol vector} \editb{$\mathbf{s}[j]$} using a per-subcarrier digital precoder $\mathbf{V}_\text{D}[j] \in \mathbb{C}^{ N_\text{RF} \times K}\editbbb{,~\forall j}$. \editbb{Then, it} appends a cyclic prefix of length \editb{$L_\text{CP} > d_{\text{max}}$}, where $d_{\text{max}}$ is the maximum delay spread of the channel, 
and subsequently applies an inverse fast Fourier transform (IFFT) operation. \editbb{Finally, the BS} employs a wide-band analog precoding matrix $\mathbf{V}_\text{RF} \in \mathbb{C}^{M \times N_\text{RF}}$. \editbb{Note that because the analog stage takes place after the IFFT module, we cannot design the analog matrix on a per subcarrier basis.} \editbl{In addition, since} the analog precoding stage is 
\editbbb{typically} implemented using \editb{a network of} phase shifters, the elements of $\mathbf{V}_\text{RF}$ \editt{must} satisfy a constant modulus constraint, i.e., $[\mathbf{V}_\text{RF}]_{mn} = e^{\imath \phi_{mn}}$\editbbb{,} where $\imath$ is the imaginary unit.

Upon receiving the signal, \editb{each user applies} an FFT operation followed by cyclic prefix removal. Mathematically, the \editb{equivalent model is given by:} 
\begin{align}
    y_k[j] &= \mathbf{h}_k^H[j] \mathbf{V}_\text{RF} \mathbf{V}_\text{D}[j] \mathbf{s}[j] 
    +  n_k[j], \nonumber \\
    &= \mathbf{h}_k^H[j] \mathbf{V}_\text{RF} \mathbf{v}_{\text{D}_{k}}[j] s_k[j] \nonumber \\ 
    &~~~~~~~+ \sum_{i \neq k} \mathbf{h}_k^H[j] \mathbf{V}_\text{RF} \mathbf{v}_{\text{D}_i}[j] s_i[j] + 
      n_\editbb{k}[j], 
\end{align}
where $y_k[j]$ 
\editbb{and $\mathbf{h}_k[j]$ are respectively the received signal and frequency-domain channel of user $k$ at subcarrier $j$}
, $\mathbf{v}_{\text{D}_k}[j]$ is the $k$-th column of $\mathbf{V}_\text{D}[j]$, and $n_k[j] \sim \mathcal{CN}(0, \sigma^2)$ is the downlink Gaussian noise. We impose a power constraint on the transmitted signal by taking $\mathbb{E} [\mathbf{s}[j] \mathbf{s}[j]^H] = \bI_K$ and $\norm{\mathbf{V}_\text{RF}\mathbf{V}_\text{D}[j]}_F^2 \leq P_\text{D}$, where $P_\text{D}$ is the per-subcarrier power budget in the downlink.

\editbb{For such a system}, the overall achievable \editb{downlink} rate for user \editb{$k$} is given by the sum of rates across subcarriers, i.e.,
\begin{equation}
    R_k = \sum_{j = 1}^{N_c} \log_2 \left(1 + \frac{\lvert \bh_k^H[j] \bV_\text{RF} \bv_{\text{D}_k}[j] \rvert^2}{ \sum_{i\not=k} \lvert \bh_k^H[j] \bV_\text{RF}\bv_{\text{D}_i}[j] \rvert^2 + \sigma^2 } \right).
\end{equation}

The main objective in this \editbb{paper} is to design the hybrid precoding matrices so as to maximize the sum rate $\sum_k R_k$. To do this, the BS 
\editb{must} obtain information about the channels. In this paper, we 
\editb{assume} that the BS has no prior knowledge of the channels, but it can acquire noisy measurements of \editbbb{the} channels through \editf{a} 
pilot 
phase.  \editt{Further, we assume that channel reciprocity \editf{holds}~\cite{smith2004reciprocity} and the \editf{downlink} CSI can be acquired in the uplink direction \editf{in a TDD operation}.} 
\begin{figure}[t]
    \centering
    \includegraphics[width=0.45\textwidth]{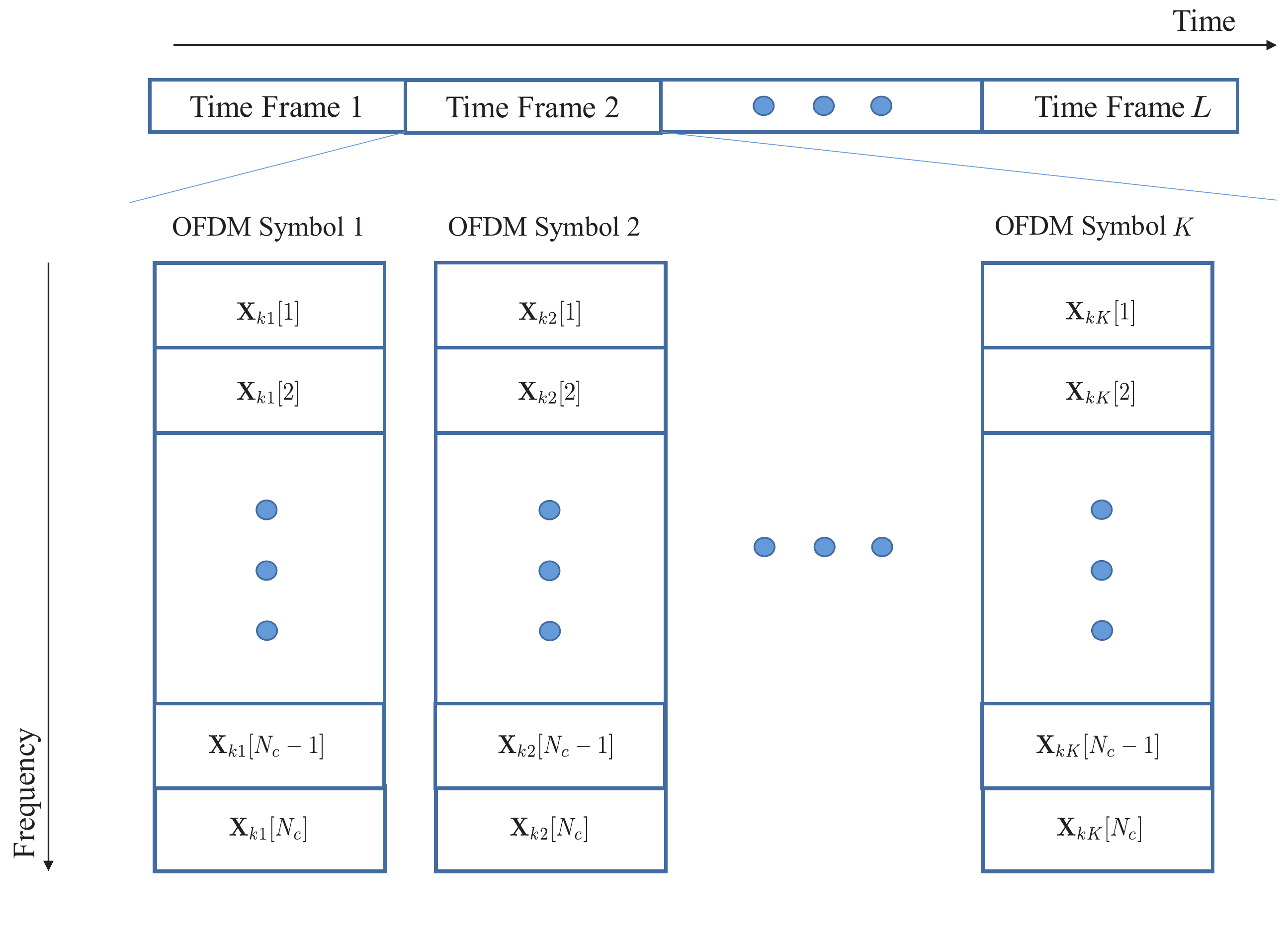}
     \caption{The \editbbb{orthogonal} uplink pilot scheme shown for the $k$-th user. In the figure, we denote the element of $\mathbf{X}[j]$ in the $m$-th row and $n$-th column by $\mathbf{X}_{mn}[j]$.}
     \label{fig:pilot_scheme}
     \vspace{-10pt}
\end{figure}
We adopt an uplink pilot transmission 
scheme that takes place over $L$ time frames each spanning $K$ OFDM symbols. In particular, during a single time frame, the users send orthogonal pilots given by the rows of the matrices $\mathbf{X}[1], \ldots, \mathbf{X}[N_c]$ over the $K$ OFDM symbols, where each pilot matrix is a $K \times K$ unitary matrix. \figurename~\ref{fig:pilot_scheme} illustrates this pilot scheme for the $k$-th user. 

Since the \editbl{number of RF chains is limited},  
the BS \editbb{senses pilot transmission in the $\ell$-th time frame using an analog sensing matrix $
\mathbf{W}_\text{RF}^{(\ell)} \in \mathbb{C}^{N_{\text{RF}} \times M}$, 
where the matrix elements satisfy 
$[\mathbf{W}^{(\ell)}_\text{RF}]_{mn} = e^{\imath \psi_{mn}}$.} After that, it right-multiplies the received signal at the $j$-th subcarrier by $\mathbf{X}[j]^H$. Due to the orthogonality of the pilots (i.e., $\mathbf{X}[j]\mathbf{X}[j]^H = \mathbf{I}_K$)
, the resulting $N_\text{RF} \times 1$ received vector for the $k$-th user in the $\ell$-th time frame is given by: 
\begin{equation}
\label{eq:y_l_ofdm}
    \tilde{\mathbf{y}}_k^{(\ell)}[j]
    = \sqrt{P_\text{U}}\mathbf{W}_{\text{RF}}^{(\ell)} \mathbf{h}_k[j] + \tilde{\mathbf{z}}_k^{(\ell)}[j],
\end{equation}
where $\tilde{\mathbf{z}}_k^{(\ell)}[j] \sim \mathcal{CN}\left(\mathbf{0}, \sigma^2 \mathbf{W}_{\text{RF}}^{(\ell)} \left(\mathbf{W}_{\text{RF}}^{(\ell)}\right)^H \right)$ is the equivalent uplink noise, and $P_\text{U}$ is the users' power budget per subcarrier in a single time frame. 
Define  $\mathbf{H}_k \triangleq  \left[ \mathbf{h}_k[1], \ldots, \mathbf{h}_k[N_c]\right]  \in\mathbb{C}^{M \times N_c}$ as the overall channel matrix for user $k$ over the entire frequency band and
\begin{equation}
    \widetilde{\mathbf{Y}}_k \triangleq \left[ \begin{matrix} \tilde{\mathbf{y}}^{(1)}_{k}[1] & \ldots & \tilde{\mathbf{y}}^{(1)}_{k}[N_c] \\ \vdots & \ddots & \vdots \\ \tilde{\mathbf{y}}^{(L)}_{k}[1] & \cdots & \tilde{\mathbf{y}}^{(L)}_{k}[N_c] \end{matrix}\right] \in \mathbb{C}^{LN_{\text{RF}} \times N_c}
\end{equation}
as the aggregate received matrix for user $k$ in $L$ time frames over the entire frequency band. Based on~\eqref{eq:y_l_ofdm}, we can write:
\begin{equation}
\label{eq:y_OFDM}
\widetilde{\mathbf{Y}}_{k} = \sqrt{P_\text{U}}\mathbf{W}_{\text{RF}} \mathbf{H}_k + 
\widetilde{\mathbf{Z}}_k, 
\quad \forall k \in \{1, \ldots, K\}
\end{equation}
where $\mathbf{W}_{\text{RF}} \editll{\triangleq} \left[ 
\left(\mathbf{W}_\text{RF}^{(1)}\right)^T, \ldots, \left(\mathbf{W}_\text{RF}^{(L)}\right)^T
\right]^T $ is the overall sensing matrix and $\widetilde{\mathbf{Z}}_k$ is a 
noise matrix whose columns are distributed as $\mathcal{CN}\left(\mathbf{0}, \sigma^2 \mathbf{A}\mathbf{A}^H \right)$ with $\mathbf{A} \triangleq \diag \left( \mathbf{W}^{(1)}_\text{RF}, \ldots, \mathbf{W}^{(L)}_\text{RF}\right)$.


\editb{For the \editbb{described} system, the hybrid precoding design problem can be stated as follows.} 
Given the noisy measurements $\{\widetilde{\mathbf{Y}}_k\}_{k = 1}^K$ \editt{obtained through the sensing matrix $\mathbf{W}_\text{RF}$} at the BS, we seek to construct the analog and digital precoding matrices according to: 
\begin{equation}
    \left(\mathbf{V}_{\text{RF}}, \mathbf{V}_{\text{D}}[1], \ldots, \mathbf{V}_{\text{D}}[N_c]\right) = \mathcal{F}\left(\widetilde{\mathbf{Y}}_1, \ldots, \widetilde{\mathbf{Y}}_K\right),
\end{equation}
so as to maximize the sum-rate expression, where $\mathcal{F}(\cdot)$ is a function that maps the received pilots into the hybrid precoding matrices. Mathematically, this can be expressed 
in terms of the following optimization problem:
\begin{align}
\label{main_problem}
\displaystyle{\Maximize_{\mathbf{W}_{\text{RF}},\mathcal{F}(\cdot)}} ~  & \sum_{k=1}^{K} \sum_{j = 1}^{N_c} \log_2 \left(1 + \frac{\lvert \bh_k^H[j] \bV_\text{RF} \bv_{\text{D}_k}[j] \rvert^2}{ \sum_{i\not=k} \lvert \bh_k^H[j] \bV_\text{RF}\bv_{\text{D}_i}[j] \rvert^2 + \sigma^2}\right) \nonumber\\ 
\text{subject to}  ~ &     \left(\mathbf{V}_{\text{RF}}, \mathbf{V}_{\text{D}}[1], \ldots, \mathbf{V}_{\text{D}}[N_c]\right) = \mathcal{F}\left(\widetilde{\mathbf{Y}}_\text{1}, \ldots,\widetilde{\mathbf{Y}}_K\right), \nonumber\\
&\left|[\mathbf{W}_\text{RF}]_{mn}\right| = 1, ~~\forall m,n, \nonumber \\
& \left|[\mathbf{V}_\text{RF}]_{m'n'}\right| = 1, ~~\forall m',n'\editll{,} \nonumber\\
& \operatorname{Tr}\left(\bV_{\text{D}}^H[j] \bV_{\text{RF}}^H \bV_{\text{RF}} \bV_{\text{D}}[j]\right) \leq P_\text{D}, ~~ \forall j.
\end{align}
Note that the overall sensing matrix \editbbb{$\mathbf{W}_\text{RF}$} \editbb{is incorporated} in the above problem as an additional optimization variable because \editbb{it} serves the critical role of summarizing the information about the user channels. 
\editbb{We remark that 
\editbbb{solving the optimization problem in~\eqref{main_problem} directly using conventional optimization based methods} is challenging due to the nonconvexity of the objective and constraints}. 
\editbl{Accordingly,} the traditional precoding schemes \editbb{seek to heuristically solve this problem} by adopting a two-step process in which the user channels are first estimated and the downlink precoding matrices are subsequently designed based on the estimated channels. 
\editb{We demonstrate in the next section} that this approach is far from optimal and it is 
advantageous to bypass the channel estimation step. 
\subsection{Channel Model}
\label{sec:chl_mdl}
\editggg{The mmWave environment is modeled as a frequency-selective channel. In particular, the channel of the $k$-th user at the $j$-th subcarrier is expressed as the $N_c$-point discrete Fourier transform (DFT) of its impulse response:}
\begin{equation}
    \editgg{\mathbf{h}_k}[j] =\sum_{n = 0}^{N_c} \editgg{\mathbf{r}_k}[n]    e^{-\imath \tfrac{2 \pi j n}{N_c}} 
    =\sum_{n = 0}^{d_\text{max}} \editgg{\mathbf{r}_k}[n]    e^{-\imath \tfrac{2 \pi j n}{N_c}},
\end{equation}
\editf{where} $\editgg{\mathbf{r}_k}[n]$ is the discrete-time \editggg{impulse response of the} channel \edit{for user $k$} \editg{at \editggg{time} index} $n$ \editf{and the second equality follows from the assumption that the channel response has a maximum delay spread $d_\text{max}$}. Further, we assume that 
\editgg{the discrete-time channel response} follows a sparse model as the sum of $L_p$ dominant paths. 
\edit{Accounting} for the effects of pulse shaping~\cite{Rodriguez2018}, \edit{we have:}
\begin{equation}
        \label{eqn:r_n}
        \editgg{\br_k[n] = \frac{1}{\sqrt{L_p}} \sum_{\ell=1}^{L_p} \alpha_{\ell, k}p_\text{rc}(nT_\text{s} - \tau_{\ell, k})\mathbf{a}_\text{t}(\theta_{\ell, k}, \phi_{\ell, k}),}
\end{equation}
where $T_s$ denotes the sampling period, $p_{\text{rc}}(\cdot)$ denotes the \editgg{normalized} raised cosine pulse-shaping filter \editgg{(i.e., $p_{\text{rc}}(0) = 1$)}, \editggg{and for simplicity $L_p$ is assumed to be the same for all users}. In addition, $\editgg{\alpha_{\ell, k}} \sim \mathcal{CN}(0, 1)$ is the complex path gain, \editgg{$\tau_{\ell, k}$} is the path delay, \editgg{$\theta_{\ell, k}$} \editbl{and} \editgg{$\phi_{\ell, k}$} are the uniformally distributed angles of departure (AoDs) in the elevation and azimuth, and $\editbb{\mathbf{a}_\text{t}}(\cdot, \cdot)$ denotes the array response vector. These channel parameters are generated independently and identically for every user. 
In addition, we assume that the BS is equipped with a two-dimensional uniform planar array (UPA) with $M_h$ and $M_v$ antennas in the horizontal and vertical directions, with $M = M_h M_v$. Moreover, the antenna separations in both directions are $\Delta_h$ and $\Delta_v$, where $\Delta_h = \Delta_v = \Delta$. The array response vector for such configuration is given by~\cite{alkhateeb2019deepmimo}:
\begin{equation}
\mathbf{a}_\editbb{\text{t}}\left(\theta, \phi\right) =   \mathbf{a}_\text{h}\left(\theta, \phi\right) \otimes \mathbf{a}_\text{v}\left(\phi\right), 
\end{equation}
where $\mathbf{a}_\text{h}(\cdot, \cdot)$ and $\mathbf{a}_\text{v}(\cdot)$ are respectively the array response vectors of a uniform linear array in the horizontal and vertical directions, i.e., {\small
\begin{align*}
    \mathbf{a}_\text{h} \left(\theta, \phi \right) &= \left[1,e^{\imath\tfrac{2\pi}{\lambda}\Delta \cos(\phi) \sin(\theta)},\ldots,e^{\imath\tfrac{2\pi}{\lambda}\Delta (M_\text{h}-1)\cos(\phi) \sin(\theta)}\right]^T, \\
    \mathbf{a}_\text{v} \left(\phi \right) &= \left[1,e^{\imath\tfrac{2\pi}{\lambda}\Delta \sin(\phi)},\ldots,e^{\imath\tfrac{2\pi}{\lambda}\Delta (M_\text{v}-1)\sin(\phi)}\right]^T,
\end{align*}}
and $\lambda$ is the wavelength.
\section{Hybrid Precoding Design for Frequency-Flat Single-Carrier Systems}
\label{sec:sc}
To \editb{fix ideas and to gain more insight}, 
we begin by considering the simpler problem of hybrid precoding design in single-carrier systems. 
For a single-carrier mmWave massive MIMO system, the hybrid precoding problem entails the design of the analog precoding matrix $\mathbf{V}_\text{RF} \in \mathbb{C}^{M \times N_\text{RF}}$ and a single digital precoding matrix $\mathbf{V}_\text{D} \in \mathbb{C}^{N_\text{RF} \times K}$. 
The uplink baseband received signal $\tilde{\mathbf{y}}_k = \left[ \left( \tilde{\mathbf{y}}_k^{(1)} \right)^T, \ldots,  \left( \tilde{\mathbf{y}}_k^{(L)} \right)^T\right]^T \in \mathbb{C}^{L N_\text{RF}}$ for each user \editbb{is given by}:
\begin{equation}
\label{eq:y_k}
    \tilde{\mathbf{y}}_k = \sqrt{P_U} \mathbf{W}_\text{RF} \mathbf{h}_k + \tilde{\mathbf{z}}_k,
\end{equation}
where \editbb{$\mathbf{h}_k$ is the channel vector and $\tilde{\mathbf{z}}_k \sim \mathcal{CN}\left(\mathbf{0}, \sigma^2 \mathbf{A} \mathbf{A}^H \right)$ is the effective uplink noise for user~$k$}. As a result, the optimization in~\eqref{main_problem} boils down to:
\begin{align}
\label{prob:sc}
\displaystyle{\Maximize_{\mathbf{W}_{\text{RF}},\mathcal{F}(\cdot)}} ~~  & \sum_{k=1}^{K} \log_2 \left(1 + \frac{\lvert \bh_k^H \bV_\text{RF} \bv_{\text{D}_k} \rvert^2}{ \sum_{i\not=k} \lvert \bh_k^H \bV_\text{RF}\bv_{\text{D}_i} \rvert^2 + \sigma^2}\right) \nonumber\\ 
\text{subject to}  ~~ &     \left(\mathbf{V}_{\text{RF}}, \mathbf{V}_{\text{D}} \right) = \mathcal{F}(\tilde{\mathbf{y}}_\text{1}, \ldots,\tilde{\mathbf{y}}_K),\nonumber\\ 
&\left|[\mathbf{W}_\text{RF}]_{mn}\right| = 1, ~~\forall m,n, \nonumber\\
& \left|[\mathbf{V}_\text{RF}]_{m'n'}\right| = 1, ~~\forall m',n'\editbl{,} \nonumber\\
& \operatorname{Tr}\left(\bV_{\text{D}}^H \bV_{\text{RF}}^H \bV_{\text{RF}} \bV_{\text{D}}\right) \leq P_\text{D}.
\end{align}
In effect, we assume $d_\text{max} = 0$. This leads to a narrow-band frequency-flat channel. \edittt{In this case, $\tau_{\ell, k} = 0$ and the channel for the $k$-th user becomes:} 
\begin{equation}
    \editgg{\mathbf{h}_{k} \triangleq \mathbf{r}_k[0]} = 
         \frac{1}{\sqrt{L_p}} \sum_{\ell=1}^{L_p} \editgg{\alpha_{\ell, k}}\mathbf{a}_\text{t}(\theta_{\ell, k}, \phi_{\ell, k}).
\end{equation}
\editrr{We now discuss the motivation behind the proposed approach.}
\subsection{\editrr{Motivation}}
\label{sec:motivation}
The hybrid precoding problem in~\eqref{prob:sc} is non-trivial to solve, \edit{because} the objective and some constraints are nonconvex. 
To tackle this problem, the conventional method is to assume 
a mathematical model of the channel and proceed to estimate the model parameters. Subsequently, the channel is reconstructed and the rate maximizing downlink precoding matrices are obtained based on \editbb{the channel reconstruction}. In the channel estimation step, this conventional paradigm \editt{needs to} introduce \editt{a} metric \editt{to} measure the distance between the true channel and \editbb{the} estimated counterpart.
The \edit{hope} is that by minimizing such \editt{a} metric, the estimated channel \editt{would} accurately approximate the true channel well enough for precoding purposes. \edit{The problem is, however, such a metric (e.g., regularized square loss for mmWave channels) is chosen for the purpose of estimating the channel parameters of the assumed model} without directly accounting for the effect of estimation errors in the subsequent precoding step. 
This implies that \edit{minimizing such a metric in} the channel estimation step 
may not exactly match the ultimate goal of maximizing the system performance (i.e., the sum rate). We argue that this possible mismatch in design metric is the main limitation of the conventional design, but is otherwise necessary to ensure that the conventional design problem is tractable. 

The advent of data-driven techniques 
shifts the paradigm toward a new possibility. Specifically, 
it is no longer necessary to adhere to the previous separation 
of channel estimation and downlink precoding, i.e., it may be beneficial to bypass the explicit CSI estimation altogether and to directly design the hybrid precoders from the baseband received pilots.
In other words, because of the power of DNN as a universal function approximator, it is now possible to pursue an end-to-end design encompassing both channel sensing/estimation and hybrid precoding.
\begin{figure}[t]
\centering
 \includegraphics[width=0.4\textwidth]{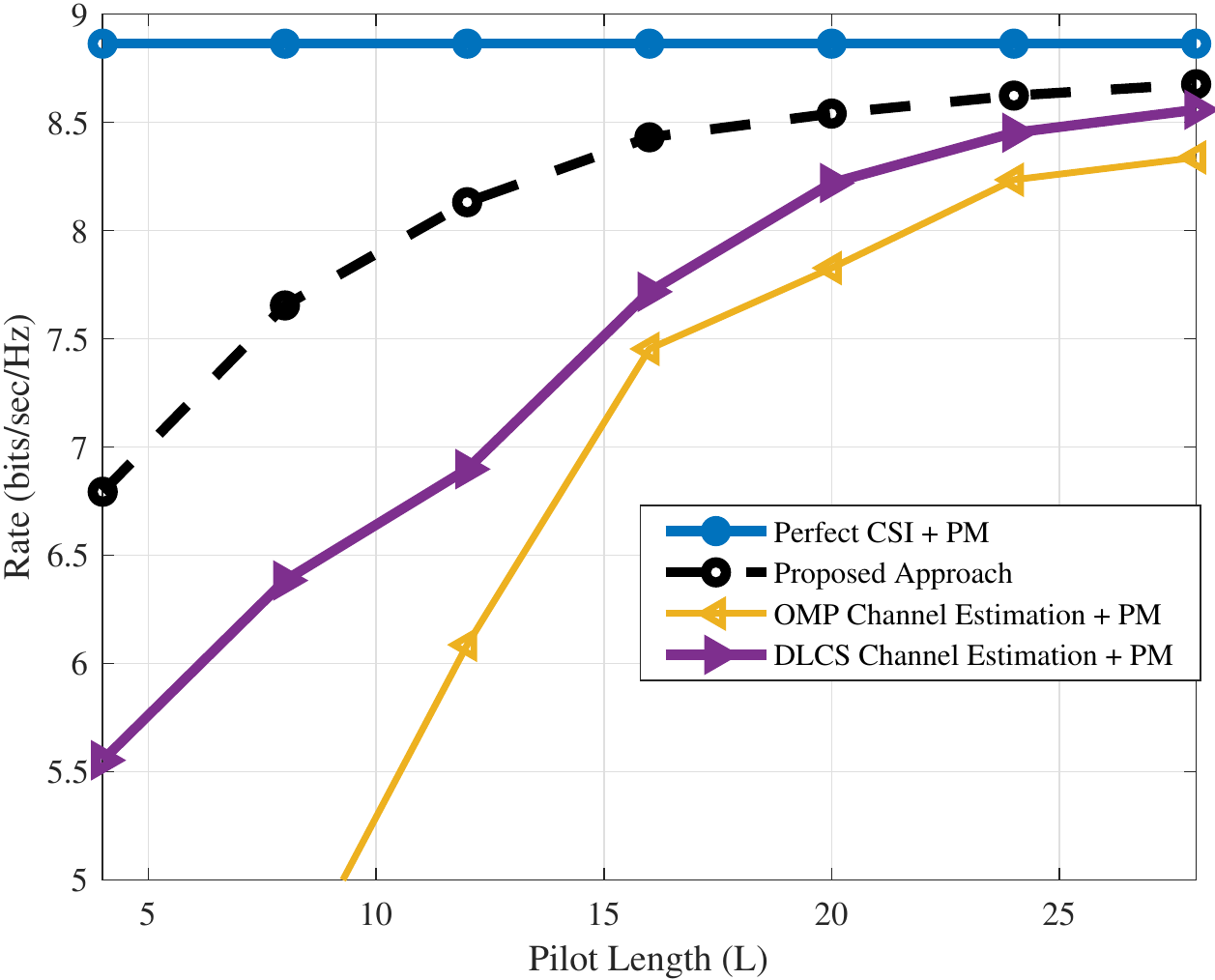}           
        \caption{Performance comparison of a single-carrier system. \editll{We set $M = 64$, $N_\text{RF} = K = 1$, and uplink/downlink SNR $= 10$dB.}}
        \label{fig:NRF1}
    \vspace{-10pt}
\end{figure}
To illustrate the preceding concepts and to quantify the impact of metric mismatch that may arise due to channel estimation, we compare the performance of different precoding schemes in the simple scenario where $K = N_\text{RF} = 1$. We choose this simple case here since the interference term vanishes \editb{and the digital beamformer reduces to a scalar whose value \editbb{can be predetermined based on the downlink power $P_\text{D}$}}. Further, the rate maximizing analog precoder has a simple optimal structure that we denote by phase matching (denoted as ``PM'' in the figures)~\cite{liang2014low}. 
This comparison is shown in~\figurename~\ref{fig:NRF1}. The proposed precoding scheme (described in Section~\ref{sec:proposed}) is a data-driven approach that maps the received pilots directly into the analog precoding vector. 
The two baseline schemes estimate the channel parameters using either the OMP~\cite{Lee2016} or a neural network~\cite{ma2020} and subsequently apply \editf{phase matching} on the estimated channels. From this figure, we can observe \editbb{sizable gain due to} bypassing channel estimation. We emphasize that such gain manifests itself when the resources for CSI acquisition are limited, e.g., when the pilot length is short. 
For sufficiently long pilot lengths, accurate channel estimation is possible and the conventional channel recovery based scheme is near optimal. 

The discussion so far highlights the advantage of directly designing the hybrid precoding matrices and advocates the use of data-driven approaches to undertake this task. However, \editbb{for the multiuser hybrid precoding problem in~\eqref{prob:sc}, a direct application of a data-driven approach is \editbbb{cumbersome}}. 
Consider a naive implementation in which we seek to learn the mapping $\mathcal{F}(\cdot)$ directly using a DNN. For simplicity,  we assume in this example that the analog sensing matrix $\mathbf{W}_{\text{RF}}$ \editbbb{is given}. Hence, the task \editbb{is reduced to} that of learning $\mathcal{F}(\cdot)$ only. Let $\mathbf{\Theta}$ be the parameter set of the DNN architecture that models this downlink precoding system, the sum rate optimization problem can now be stated as follows:
\begin{align}
\displaystyle{\minimize_{\mathbf{\Theta}}} ~  & \editff{\mathbb{E}_{\mathbf{H}, \widetilde{\mathbf{Z}}}
\left[\sum_{k=1}^{K} \log_2 \left(1 + \frac{\lvert \bh_k^H \bV_\text{RF} \bv_{\text{D}_k} \rvert^2}{ \sum_{i\not=k} \lvert \bh_k^H \bV_\text{RF}\bv_{\text{D}_i} \rvert^2 + \sigma^2}\right) \right]} \\ 
\text{subject to}  ~ &     \left(\mathbf{V}_{\text{RF}}, \mathbf{V}_{\text{D}}\right) = \mathcal{F}_{\text{NN}}\left(\tilde{\mathbf{y}}_1, \ldots, \tilde{\mathbf{y}}_K; \mathbf{\Theta}\right), \nonumber 
\end{align}
where $\mathcal{F}_{\text{NN}}(~\cdot~ ; \mathbf{\Theta})$ represents the input-output relationship of the DNN and the expectation is taken over both the channel $\mathbf{H} = \left[\mathbf{h}_1, \ldots, \mathbf{h}_K\right]$ and the uplink noise $\widetilde{\mathbf{Z}} = \left[\tilde{\mathbf{z}}_1, \ldots, \tilde{\mathbf{z}}_K\right]$ in the sensing stage due to the dependence of $\mathbf{V}_\text{RF}$ and $\mathbf{V}_\text{D}$ on the noise realizations through $\tilde{\mathbf{y}}_1, \ldots, \tilde{\mathbf{y}}_K$. \editbb{\editll{We} can learn the \editll{DNN} parameters by \editff{replacing the expectation with the empirical average over some training set $\mathcal{T}$ and} subsequently employing a stochastic gradient descent (SGD) algorithm.}

While it is easy to see that this method  bypasses the channel\editbb{-}recovery step and avoids the 
\editbbb{limitations of} the traditional precoding scheme, it turns out that such an approach 
suffers from a number of drawbacks. 
First, the training process is severely hindered by complexity \editr{due to} 
the multiplicative interaction between the analog and digital precoding matrices. 
\editbbb{Second, the input and output dimensions of this DNN depend on the number of users. 
%
\editr{This} implies that 
a different DNN \edit{needs to be trained} 
for systems with different number of users}. For these reasons, this naive approach is difficult to implement in practice and is not generalizable. Thus, \editt{an} alternative learning strategy \editbbb{is needed.} 
\subsection{Proposed Precoding Design}
\label{sec:proposed}
This section develops the proposed \editbbb{semi-direct} data-driven scheme that overcomes the limitation of \editbbb{channel recovery} based approaches while enabling simpler training and better generalizability than \editbbb{the aforementioned naive application of the DNN}. The key idea behind the proposed approach is two-fold. First, we decouple the design of the analog and digital precoding stages and \editbbb{learn a direct mapping for the analog precoder only. 
This reduces the training complexity of the DNN as compared to the one that simultaneously designs the hybrid precoders.}  
Second, we decompose the overall DNN architecture into several \emph{per-user DNNs}, each outputting one column of $\mathbf{V}_\text{RF}$. \editrev{Further, we tie the weights of all the per-user DNNs together and show that training one per-user DNN using the training data of all users is sufficient to learn a universal mapping for all users. This allows us to train one per-user DNN, but employ any number of copies of the trained DNN to serve an arbitrary number of users during the actual operation of the system.} 

\begin{figure}[t]
        \centering
        \includegraphics[width=0.5\textwidth]{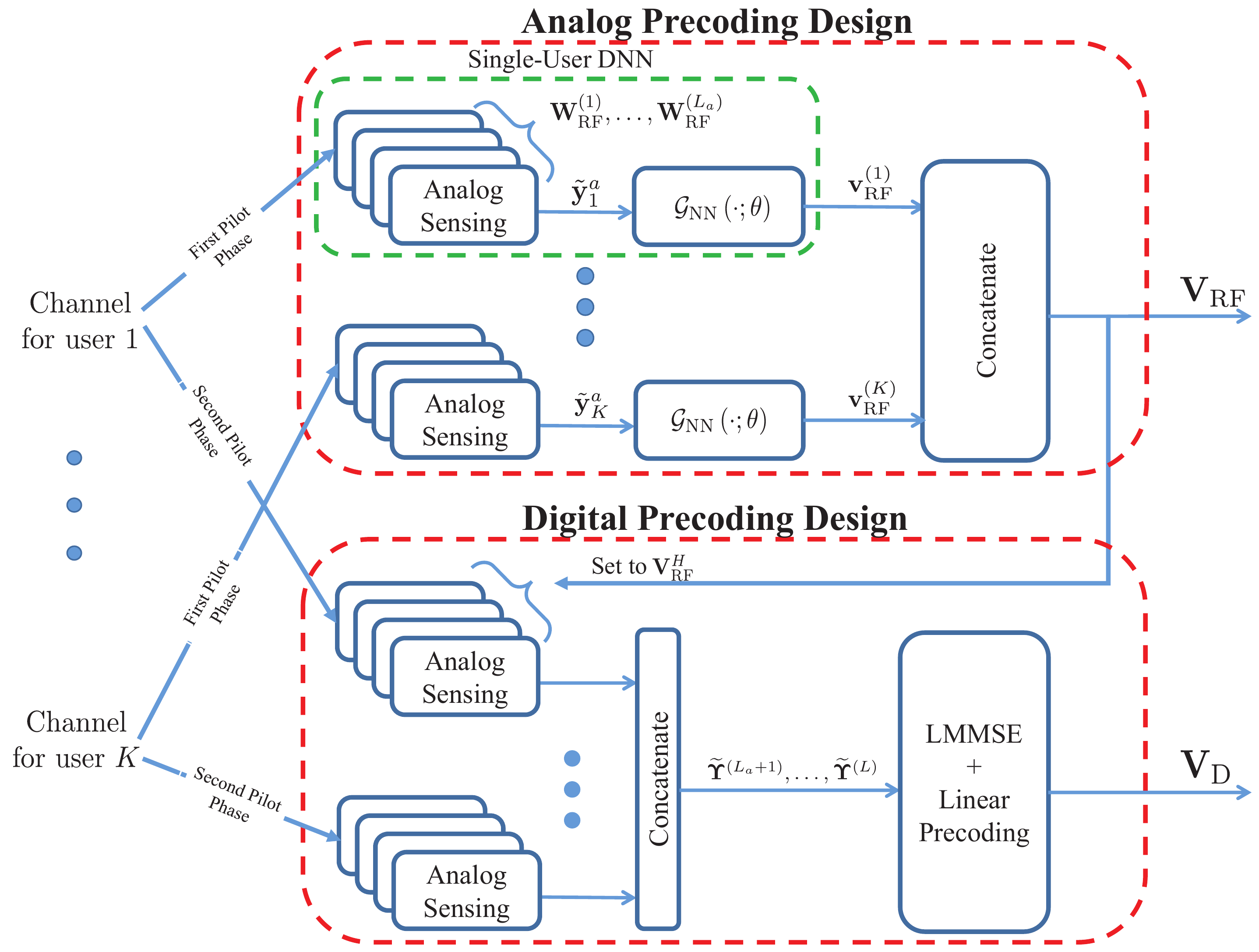}
        \caption{Block diagram of the proposed approach for designing the hybrid precoding matrices in a single-carrier \editbl{TDD} system.}
           \vspace{-10pt}
        \label{fig:blk_diag}
\end{figure}

In order to decouple the design of $\mathbf{V}_{\text{RF}}$ and $\mathbf{V}_{\text{D}}$, we propose to
split the overall \editrev{pilot phase} of $L$ frames into \editrev{two separate phases of lengths $L_a$ and $L_d$, where $L = L_a + L_d$. To design the analog precoder, we map the pilots received in the first pilot phase into the analog precoding matrix using a DNN. Then, we fix the analog beamformers and initiate the second phase of pilot transmission. The pilots received in the second phase are then used to estimate the low-dimensional equivalent channel and to design the digital precoding matrix using a conventional approach that requires no further training.} 
The block diagram of the proposed scheme is shown in \figurename~\ref{fig:blk_diag}. 
\subsection{\editrev{Analog Precoding Design}}
\label{sec:analog_part_design}
\editbbb{We assume that} $K = N_{\text{RF}}$ so that $\mathbf{V}_{\text{RF}}$ has $K$ columns. This assumption is made in most hybrid precoding works \cite{liang2014low, ma2020} 
due to \editbbb{the fact that a fully loaded system maximizes the overall multiplexing gain}. 
Let us denote the received pilot from the $k$-th user in the \editrev{first pilot phase} 
by $\tilde{\mathbf{y}}^a_k
\triangleq \left[ \left(\tilde{\mathbf{y}}_{k}^{(1)}\right)^T,  \ldots ,  \left(\tilde{\mathbf{y}}_{k}^{(L_a)}\right)^T\right]^T$, then \editbl{using}~\eqref{eq:y_k}, we may write:
\begin{equation}\label{eqn:y_a}
    \tilde{\mathbf{y}}_k^{a} = \sqrt{P_\text{U}} \mathbf{W}_{\text{RF}}^a \mathbf{h}_k + \tilde{\mathbf{z}}_k^{a},
\end{equation}
where $\mathbf{W}_{\text{RF}}^a \triangleq \left[\begin{matrix} \left(\mathbf{W}_{\text{RF}}^{(1)}\right)^T,  \ldots,   \left(\mathbf{W}_{\text{RF}}^{(L_a)}\right)^T \end{matrix}\right]^T$ and \editbbb{$\tilde{\mathbf{z}}^a_k \sim \mathcal{CN} \left(\mathbf{0}, \sigma^2 \mathbf{B}\mathbf{B}^H \right)$ is the corresponding uplink noise with $\mathbf{B} = \diag\left(\mathbf{W}_\text{RF}^{(1)}, \ldots, \mathbf{W}_\text{RF}^{(L_a)} \right)$.} We seek to determine the analog precoder directly from the received pilots in~\eqref{eqn:y_a} using a DNN on a per-user basis. To accomplish this, we \editbbb{decompose} 
the overall DNN architecture 
into $K$ non-interconnected branches. 
In the $k$-th branch, we have a single-user DNN \editll{(SU-DNN)}, \editt{with parameters $\mathbf{\Theta}_k$,} that determines the $k$-th analog precoding vector $\mathbf{v}_{\text{RF}}^{(k)}$ from $\tilde{\mathbf{y}}_k^a$ using the following direct \editbbb{mapping}:
\begin{equation}
    \label{eq:v_rf_k}
    \mathbf{v}^{(k)}_{\text{RF}} = \mathcal{G}_{\text{NN}}(\tilde{\mathbf{y}}^a_k; \mathbf{\Theta}_k), \qquad{\forall k}.
\end{equation}
\editrev{We make the design choice to construct the analog precoding vector on a per-user basis in order to simplify the training process and to further ensure that the proposed scheme can generalize to systems with any number of users.} \editrev{Based on~\eqref{eq:v_rf_k}, the} goal is to find the DNN parameters in each branch $\mathbf{\Theta}_k$
and the sensing matrix $\mathbf{W}_{\text{RF}}^a$ so that some average loss function is minimized:
\begin{align}
\label{eq:analog_opt}
\displaystyle{\minimize_{\mathbf{W}^a_{\text{RF}}, \{\mathbf{\Theta}_k\}_{k = 1}^K}} ~~  & \editff{\mathbb{E}_\editrev{\mathbf{H}, \widetilde{\mathbf{Z}}^a}
\left[ \mathcal{L}(\mathbf{V}_{\text{RF}}) \right]}\\ 
\text{subject to}  ~~~ &         \mathbf{v}^{(k)}_{\text{RF}} = \mathcal{G}_{\text{NN}}(\sqrt{P_\text{U}} \mathbf{W}_{\text{RF}}^a \mathbf{h}_k + \tilde{\mathbf{z}}_k^{a}; \mathbf{\Theta}_k), \quad{\forall k}. \nonumber
\end{align}
where $\widetilde{\mathbf{Z}}^a \triangleq \left[\tilde{\mathbf{z}}^a_1, \ldots,  \tilde{\mathbf{z}}^a_K\right] \in \mathbb{C}^{L N_\text{RF} \times K}$. Strictly speaking, since \editbbb{the ultimate goal is 
sum rate maximization,} 
the loss function should be defined in terms of the 
maximum sum rate optimized over all choices of the digital precoder. In other words, for a given $\mathbf{V}_{\text{RF}}$, the loss function should be the negative of the optimal value obtained from solving the following optimization problem:
\begin{align}
\label{prob:intractable}
&\mathcal{L}^\text{opt} \left( \mathbf{V}_\text{RF} \right) = \nonumber \\ &- \displaystyle{\Maximize_{\small \mathbf{V}_\text{D} : \left\|  \bV_{\text{RF}} \bV_{\text{D}}\right\|_F^2 \leq P_\text{D}}} 
\sum_{k=1}^{K} \log_2 \left(1 + \frac{\lvert \bh_k^H \bV_\text{RF} \bv_{\text{D}_k} \rvert^2}{ \sum_{i\not=k} \lvert \bh_k^H \bV_\text{RF}\bv_{\text{D}_i} \rvert^2 + \sigma^2}\right) 
\end{align}
However, this problem does not have a closed-form solution; it is nonconvex so even a numerical solution is nontrivial. \editrev{Instead, we obtain a simplified closed-form approximation to $\mathcal{L}^\text{opt}\left(\mathbf{V}_\text{RF}\right)$ based on the following steps: (i) we ignore the interference term in the denominator  of the objective of~\eqref{prob:intractable}, and (ii) we simplify the resulting expression into a per-user form.

We remark that ignoring the interference in the design of analog precoder is a common heuristic in hybrid precoding literature, e.g., see~\cite{ma2020, liang2014low}. This is because 
the subsequent digital precoder is typically designed to alleviate the effect of interference, e.g., using zero forcing
(ZF). With the interference term removed, the optimization problem then becomes:
\begin{equation}
\label{prob:loss_1}
\displaystyle{\Maximize_{\mathbf{V}_\text{D} : \left\|  \bV_{\text{RF}} \bV_{\text{D}}\right\|_F^2 \leq P_\text{D}}} 
~~  \sum_{k=1}^{K} \log_2\left( 1 + \frac{\left| \mathbf{h}_k^H \mathbf{V}_\text{RF} \mathbf{v}_{D_k}\right|^2  }{\sigma^2} \right).
\end{equation}
The solution of this optimization problem has a closed-form structure, given by maximal ratio transmission $\mathbf{V}^*_\text{D} = \mathbf{V}
_\text{RF}^H \mathbf{H} \mathbf{D}$, where $\mathbf{D} = \diag{\left(\tfrac{\sqrt{d_1}}{\|\mathbf{V}_\text{RF}^H \mathbf{h}_1\|}, \ldots, \tfrac{\sqrt{d_K}}{\|\mathbf{V}_\text{RF}^H \mathbf{h}_K\|} \right)}$ and $d_k$ is the power allocated to the $k$-th user. With this structure of~$\mathbf{V}_\text{D}$, we approximate the power constraint as follows:
\begin{align}
    \label{eq:power_constraint}
    \Tr{\left({\mathbf{V}^{*H}_\text{D}} \mathbf{V}_\text{RF}^H \mathbf{V}_\text{RF} \mathbf{V}^*_D \right)} &\approx M  \Tr{\left(\mathbf{V}^{*H}_\text{D} \mathbf{V}^*_D \right)} \nonumber \\ &
    = M \sum_{k = 1}^K d_k \leq P_\text{D},
\end{align}
where the approximation $\mathbf{V}_\text{RF}^H \mathbf{V}_\text{RF} \approx M \mathbf{I}$
follows from Lemma 1 in~\cite{SohrabiOFDM}. For simplicity, we adopt an equal power allocation, i.e., $d_1 = ... = d_k = \tfrac{P_\text{D}}{KM}$, which is near optimal at high SNR. In this case, the loss function becomes:
\begin{equation}
\label{eq:loss_before_approx}
\tilde{\mathcal{L}}\left( \mathbf{V}_\text{RF}\right) = - \sum_{k = 1}^K \log_2 \left( 1 + \frac{P_\text{D}}{MK\sigma^2} \| \mathbf{V}_\text{RF}^H \mathbf{h}_k \|^2 \right)    . 
\end{equation}

We can further simplify~\eqref{eq:loss_before_approx} by noting that 
$\mathbf{v}_\text{RF}^{(k)}$ is a function of $\mathbf{h}_k$ only, according to~\eqref{eq:v_rf_k}. Assuming that the user channels are independent, we note $\mathbf{h}_{k'}^H \mathbf{v}_\text{RF}^{(k)} \sim \mathcal{CN}(\mathbf{0}, M)$ whenever $k' \neq k$ since the elements of $\mathbf{v}_\text{RF}^{(k)}$ act as random phase rotations for the elements of $\mathbf{h}_k$.
It then follows that $\tfrac{1}{M} |\mathbf{h}_{k'}^H \mathbf{v}_\text{RF}^{(k)}|^2 \sim \text{exp}(1)$, where $\text{exp}(\lambda)$ is an exponential distribution with parameter $\lambda$. On the other hand, the distribution of $\tfrac{1}{M} |\mathbf{h}_{k}^H \mathbf{v}_\text{RF}^{(k)}|^2$ 
can be described in terms of the $\text{SNR}$ distribution of equal-gain combining, which 
is known to have drastically improved statistics with increasing $M$~\cite{brennan1959}. 
Hence, in the massive MIMO regime where $M$ is large, we can approximate $\tfrac{1}{M} \| \mathbf{V}_\text{RF}^H \mathbf{h}_k \|^2 = \tfrac{1}{M} \sum_j | \mathbf{h}_k^H \mathbf{v}_\text{RF}^{(j)}|^2 \approx \tfrac{1}{M} | \mathbf{h}_k^H \mathbf{v}_\text{RF}^{(k)}|^2$. As a result, the loss function becomes:
\begin{equation}
\label{eq:loss_function}
\mathcal{L}\left( \mathbf{V}_\text{RF}\right) = - \sum_{k = 1}^K \log_2 \left( 1 + \frac{P_D}{MK\sigma^2} | \mathbf{h}_k^H \mathbf{v}_\text{RF}^{(k)}|^2 \right).
\end{equation}  
We propose to train the DNN using~\eqref{eq:loss_function}.  

\editrev{We remark that the loss function~\eqref{eq:loss_function} bears some similarity to the metric of~\cite{ma2020} for analog precoding design. Specifically, both metrics encourage the analog precoder phase shifters to match the channel phases. The key difference is that we propose to construct the analog precoder directly from the received pilots. In contrast, the work~\cite{ma2020} estimates the channel first and uses this estimate to determine the analog precoder. The limitation of separating channel estimation and precoding modules has already been discussed in Section~\ref{sec:motivation}. In particular, most channel estimation schemes 
aim to estimate both the channel gains and phases. In contrast, the loss function~\eqref{eq:loss_function} suggests that only the channel phases are important in so far as the analog precoding design is considered, but not the channel gains. Hence, it can be seen that bypassing channel estimation has the potential to yield a better analog precoder in the short-pilot regime. This can be seen as the main advantage of the proposed design as compared to~\cite{ma2020}.} 

In terms of training, the key benefit of~\eqref{eq:loss_function} is that it} consists of $K$ independent terms each of which is a function of a single analog precoding vector 
$\mathbf{v}_{\text{RF}}^{(k)}$. 
\editrev{
In other words, the objective of the optimization problem~\eqref{eq:analog_opt} decouples over the users, i.e.,
\begin{align}
\label{eq:simp}
\mathbb{E}&_{\mathbf{H}, \tilde{\mathbf{Z}}^a}
\left[ \mathcal{L}(\mathbf{V}_{\text{RF}}) \right]
=\nonumber \\
& - \sum_{k = 1}^K \mathbb{E}_{\mathbf{h}_k, \tilde{\mathbf{z}}^{a}_k} \left[ \log_2 \left(1 +\frac{P_\text{D}}{M K \sigma^2}\left| \mathbf{h}_k^H \mathbf{v}_\text{RF}^{(k)}
\right|^2
\right) \right].
\end{align} 
To further simplify training, we propose to tie the parameters of the per-user DNNs together, i.e., set $\mathbf{\Theta}_1 = \ldots = \mathbf{\Theta}_K = \mathbf{\Theta}$. In this case, all SU-DNNs~\eqref{eq:v_rf_k} become identical. As a result, the optimization problem becomes:
\begin{align}
\label{eq:Single_branch_optimization}
\displaystyle{\minimize_{\mathbf{W}^a_\text{RF},~\mathbf{\Theta}}} ~~  & - \mathbb{E}_{\mathbf{h}, \tilde{\mathbf{z}}^a}
\left[ \log_2\left(1 + \frac{P_\text{D}}{M K \sigma^2} \left\lvert \bh^H \mathbf{v}_{\text{RF}} \right\rvert^2 \right) \right]\\ 
\text{subject to}  ~~ &         \mathbf{v}_{\text{RF}} = \mathcal{G}_{\text{NN}}(\sqrt{P_U} \mathbf{W}_\text{RF}^a \mathbf{h} + \tilde{\mathbf{z}}^a; \mathbf{\Theta}), \nonumber
\end{align}
where we define a new 
channel vector $\mathbf{h}$ to have a mixture distribution $\tfrac{1}{K} \sum_k f_{\mathbf{h}_k}\left(\mathbf{h}\right)$ with $f_{\mathbf{h}_k}(\cdot)$ denoting the distribution of the $k$-th user channel and $\tilde{\mathbf{z}}^a$ and $\mathbf{v}_\text{RF}$ are defined similarly.

We interpret~\eqref{eq:Single_branch_optimization} as learning the parameters $\{\mathbf{W}_\text{RF}^a,  \mathbf{\Theta}\}$ of  ``one" SU-DNN whose input is given by the channel plus noise and output is $\mathbf{v}_\text{RF}$. In effect, the training of the overall DNN for $K$ users can be performed by training a single SU-DNN using a training set whose samples are drawn from the mixture distribution. 
To construct this training set, 
we treat a batch of $K$-user channel realizations 
$\mathcal{B} = \{\mathbf{H}^{(1)}, \ldots, \mathbf{H}^{\left(B\right)}\}$ of size $B$ 
as a new batch $\tilde{\mathcal{B}}= \{\mathbf{h}^{(1)}, \ldots, \mathbf{h}^{(BK)} \}$ whose elements are taken columns-wise from all $\mathbf{H}^{(i)} \in \mathcal{B}$ with random shuffling. 
}

The main advantage of learning a common set of parameters in~\eqref{eq:Single_branch_optimization} is that it allows us to train one SU-DNN based on the samples of all possible realizations of user channels, then to deploy copies of the trained SU-DNN for systems with an arbitrary number of users during the actual operation. This ensures that the proposed approach does not require retraining when the number of users changes. We emphasize here that 
this method is applicable even if the users experience different channel distributions and different noise levels. 

\subsection{\editll{SU-DNN} Architecture}
\begin{figure*}[!t]
        \centering
        \includegraphics[width=0.7\textwidth]{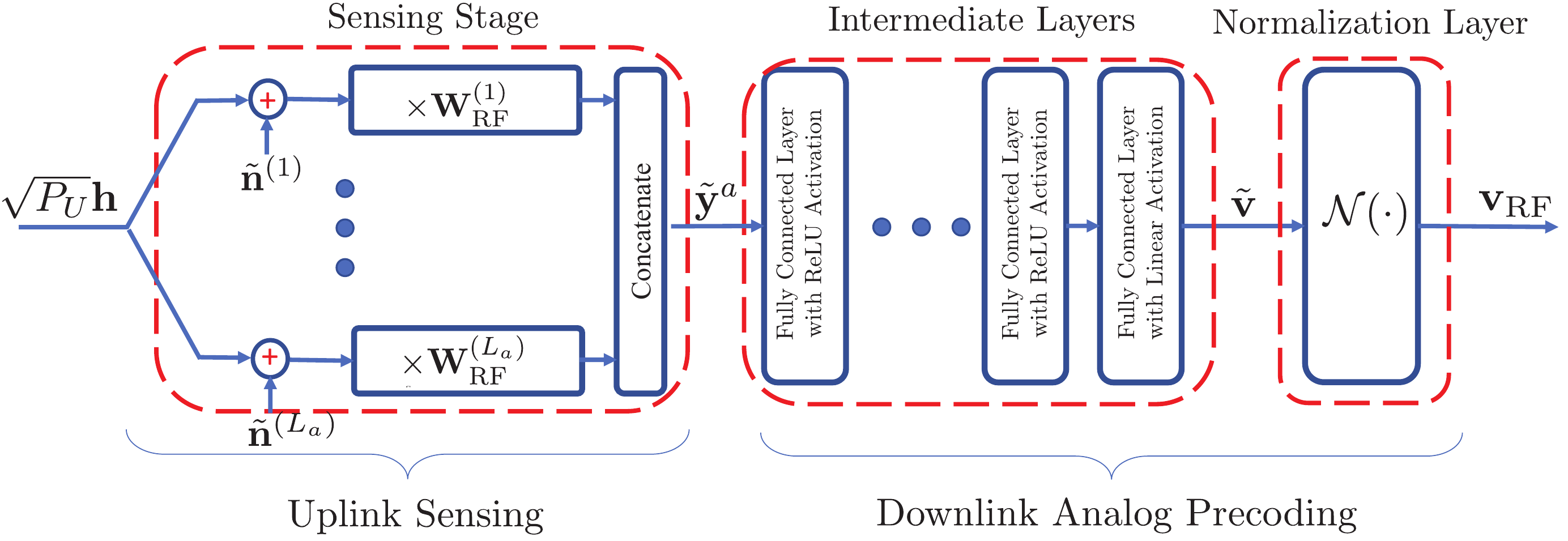}
        \caption{\editll{T}he proposed \editll{SU-DNN} for uplink sensing and downlink analog precoding design in single-carrier systems.}
             \vspace{-10pt}
        \label{fig:DNN_SC}
\end{figure*}
The proposed \editll{SU-DNN} that learns the analog sensing matrix $\mathbf{W}_{\text{RF}}^a$ and the parameter set $\mathbf{\Theta}$ is depicted in \figurename~\ref{fig:DNN_SC}.  This architecture consists of a sensing stage, a set of intermediate layers, and an entry-wise normalization layer. 
\subsubsection{Sensing stage}
\editrev{The sensing stage is a trainable stage that models the sensing operation:
\begin{equation}\label{eqn:y_a_nok}
\tilde{\mathbf{y}}^{(\ell)} = \sqrt{P_\text{U}} \mathbf{W}_{\text{RF}}^{(\ell)} \mathbf{h} + \tilde{\mathbf{z}}^{(\ell)}, \quad \ell \in \{1, \ldots, L_a\},
\end{equation}
where $\tilde{\mathbf{z}}^{(\ell)} \sim \mathcal{CN} \left(\mathbf{0}, \mathbf{W}_{\text{RF}}^{(\ell)} \left(\mathbf{W}_{\text{RF}}^{(\ell)}\right)^H \right)$. This can be implemented by expressing $\tilde{\mathbf{z}}^{(\ell)} = \mathbf{W}_{\text{RF}}^{(\ell)} \tilde{\mathbf{n}}^{(\ell)}$ for a noise vector $\tilde{\mathbf{n}}^{(\ell)} \sim \mathcal{CN} \left(\mathbf{0}, \mathbf{I}_M \right)$  and using a fully-connected (FC) linear layer with input as the channel vector plus noise (i.e., $\sqrt{P_U} \mathbf{h} + \tilde{\mathbf{n}}^{(\ell)}$), output as $\tilde{\mathbf{y}}^{(\ell)}$, and trainable weights given by the sensing matrix $\mathbf{W}_\text{RF}^{(\ell)}$. The elements of $\mathbf{W}_\text{RF}^{(\ell)}$ are restricted to have a unit modulus. To enforce such constraint on these elements, we take the trainable parameters to be the phases of $[\mathbf{W}_\text{RF}^{(\ell)}]_{ij}$.
By using $L_a$ such FC layers in parallel, we model the sensing operation~\eqref{eqn:y_a_nok}. The different outputs $\tilde{\mathbf{y}}^{(1)}, \ldots, \tilde{\mathbf{y}}^{(L_a)}$ are then concatenated to form $\tilde{\mathbf{y}}^a$}.
\subsubsection{Intermediate Layers} \editrev{The intermediate stage is a network of layers that, together with the normalization layer in~(26), models the direct map from the received pilot to the precoding vector, (i.e., $\mathcal{G}_\text{NN}\left( \cdot ; \mathbf{\Theta} \right)$). In other words, this stage combines implicit channel estimation with analog precoding.} To implement this stage, we consider a cascade of FC layers that model the direct mapping from the received pilots $\tilde{\mathbf{y}}^a$ to the unnormalized precoding vector~$\tilde{\mathbf{v}}$. The input-output relationship in this case is given by:
\begin{equation*}\label{eqn:intr_layer}
    \tilde{\mathbf{v}} = \mathbf{W}_R\mathcal{R}\left(\ldots \mathcal{R}\left(\mathbf{W}_1 \tilde{\mathbf{y}}^a + \mathbf{b}_1\right) \ldots \right) + \mathbf{b}_R,
\end{equation*}
where $R$ is the number of intermediate layers, $\mathcal{R}(\cdot) = \max(\cdot, 0)$ is the ReLU activation, which is applied to all layers except the last one, and $\mathbf{\Theta} = \{\mathbf{W}_r, \mathbf{b}_r\}_{r = 1}^R$ is the set of trainable parameters. 
\subsubsection{Normalization Layer}
The output of the intermediate stage cannot be 
taken as the analog precoding vector since the elements of $\tilde{\mathbf{v}}$ do not satisfy the unit modulus constraint. The purpose of \editbbb{the normalization} layer is to enforce this constraint on the elements of the output vector. This is accomplished by applying the map $\mathcal{N}\left(x\right) \triangleq \tfrac{x}{|x|}$ component-wise on the elements of  $\tilde{\mathbf{v}}$, i.e.,
\begin{equation}
\label{eq:norm}
    \left[\mathbf{v}_{\text{RF}}\right]_i = \mathcal{N}\left(\left[\tilde{\mathbf{v}}\right]_i \right), \quad \forall i.
\end{equation}
In our prior work \cite{attiahdeep2020}, we propose \editbbb{to output the phases of the analog precoding vector $\mathbf{\Phi}$ in the intermediate stage 
and apply the component-wise exponential map $\left[\mathbf{v}_\text{RF}\right]_{i} = e^{\imath [\mathbf{\Phi}]_{i}}$ in the subsequent layer to ensure that the analog constraint is met.} 
However, our numerical experiments \editbbb{indicate} that the normalization method\editr{~\eqref{eq:norm}} can achieve better performance. 

\subsubsection{Training and Post-Training}
The training  of \editbbb{the proposed DNN is performed offline in an unsupervised fashion to minimize the average loss function in~\eqref{eq:loss_function}}.  
After training, 
we use $K$ copies of the \editll{SU-DNN}s for operation wherein the sensing stage is used as the analog sensing matrix in the uplink pilot phase and the received pilot signal is fed into the intermediate layers to produce the downlink analog precoder.

Finally, we remark that the analog precoding and sensing stages 
discussed herein are assumed to be of infinite-resolution. In practice, the analog stage is often implemented using finite-resolution phase shifters, which implies that the phases 
take on values over a discrete set. A particular case of interest is when the elements have the form $[\mathbf{U}_{\text{RF}}]_{mn} = e^{\imath \tfrac{2 \pi q}{Q}},$ 
where $\mathbf{U}_\text{RF} \in \{\mathbf{V}_\text{RF}, \mathbf{W}_\text{RF}\}$ and $Q$ is an integer with $q = 0, \ldots, Q - 1$. To produce analog precoding/sensing matrices adhering to this restriction, we use the previous DNN architecture to determine the matrices with unrestricted phases 
then round the phases to the nearest value in the discrete set.
\subsection{\editrev{Digital Precoding Design}} 
In the digital \editrev{precoding design}, we seek to design 
the digital precoder $\mathbf{V}_\text{D}$\editbbb{, given a predetermined $\mathbf{V}_{\text{RF}}$ and using a second pilot phase of length $L_d$ time frames.} 
We propose to construct the  digital precoding using the traditional approach that separates the channel estimation and precoding modules \editll{since the performance loss due to the metric mismatch is negligible for low-dimensional channels.} 

\editbbb{For a fixed analog precoder, the low-dimensional equivalent channel seen by the digital precoder is 
$\mathbf{H}_{\text{eq}} \triangleq \mathbf{V}_\text{RF}^H \mathbf{H}.$} \editrev{To estimate the equivalent low-dimensional channel $\mathbf{H}_\text{eq}$, we use the second pilot transmission phase, where 
the received pilot matrix $\widetilde{\mathbf{\Upsilon}}^{(\ell)} \in \mathbb{C}^{M \times K}$ for all $K$ users in the $\ell$-th time frame is given by:
\begin{equation}
    \label{eq:dig_y_l2}
    \widetilde{\mathbf{\Upsilon}}^{(\ell)} \triangleq \left[ \tilde{\mathbf{y}}^{(\ell)}_1, \ldots, \tilde{\mathbf{y}}^{(\ell)}_K \right] = \sqrt{P_\text{U}} 
    \mathbf{W}_\text{RF}^{(\ell)}\mathbf{H}
    + \widetilde{\mathbf{Z}}^{(\ell)},  
\end{equation}
\editfi{where $\ell = L_a + 1, \ldots, L$.} Here, $\mathbf{W}_\text{RF}^{(L_a)}, \ldots, \mathbf{W}_\text{RF}^{(L)}$ represent the sensing matrices that can be designed. We enforce the choice:
\begin{equation}
\label{eq:wrf_vrf}
\mathbf{W}_{\text{RF}}^{(\ell)} = \mathbf{V}_{\text{RF}}^H, \quad  \ell = L_a + 1, \ldots, L,
\end{equation}
so as to make the useful signal term in~\eqref{eq:dig_y_l2} identical to $\mathbf{H}_\text{eq}$}. In particular, this choice transforms the end-to-end massive MIMO system into a fully digital low-dimensional MIMO system whose channel is $\mathbf{H}_\text{eq}$. \editf{Indeed, by~\eqref{eq:wrf_vrf},} the received pilots can now be expressed as:
\begin{equation}
\label{eq:dig_y_l}
\widetilde{\mathbf{\Upsilon}}^{(\ell)}  = \sqrt{P_\text{U}}\overbrace{\mathbf{V}^H_{\text{RF}} \mathbf{H}}^{\mathbf{H}_{\text{eq}}}  + \widetilde{\mathbf{Z}}^{(\ell)},  \quad \ell = L_a + 1, \ldots, L,
\end{equation}
where~\eqref{eq:dig_y_l} can now be regarded as repeated transmissions of the pilots through the equivalent channel. As a result, the estimate $\hat{\mathbf{H}}_\text{eq}$ can be determined directly from the received pilots $\widetilde{\mathbf{\Upsilon}}^{(L_a + 1)}, \ldots, \widetilde{\mathbf{\Upsilon}}^{(L)}$, e.g., using the traditional linear minimum mean squared error (LMMSE). 
Assuming \edit{that} \editbbb{the channels across the different antennas are uncorrelated} 
(i.e., $\mathbb{E} [\mathbf{h}_k \mathbf{h}_k^H ] = \mathbf{I}_M, \forall k$), the LMMSE estimator for the equivalent channel is given by:
\begin{equation}\label{eq:mmse}
     \hat{\mathbf{H}}_{\text{eq}} =  \frac{\sqrt{P_\text{U}}}{P_\text{U} L_d + \sigma^2} 
     \sum_{\ell = L_a + 1}^{L} \widetilde{\mathbf{\Upsilon}}^{(\ell)}. 
\end{equation}

Having estimated $\mathbf{H}_\text{eq}$, we proceed to determine $\mathbf{V}_{\text{D}}$ using conventional linear precoding \editbbb{schemes}. Two common choices \editll{for digital precoding} design are the \editbbb{\editrr{ZF} and the iterative weighted minimum mean squared error} (WMMSE) technique~\cite{shi2011iteratively}. For ZF, the digital precoder \editll{is} $
\mathbf{V}_{\text{D}} =  \hat{\bH}_{\text{eq}} (\hat{\bH}_{\text{eq}}^H \hat{\bH}_{\text{eq}})^{-1} \mathbf{D}_{\text{ZF}}$,
where $\mathbf{D}_{\text{ZF}}$ is \editll{the power allocation} diagonal matrix. Despite its simplicity and its ability to mitigate interference, ZF suffers from noise enhancement. \editll{The} WMMSE approach is an iterative strategy that reduces the combined effect of noise and interference. The WMMSE procedure for digital precoding design \editbbb{\editr{in} hybrid systems} 
is summarized in\editbbb{~\cite{SohrabiOFDM}}.
\editrev{\subsection{Pilot Allocation}
\label{sec:allocation}
There is a natural trade-off between the number of pilots allocated to the analog precoding design and those allocated to the digital precoding design. In particular, for a fixed $L$, allocating more pilots to the analog precoding design ensures a good analog precoder but may hurt the design of the digital precoder and vice versa. In general, the numerical simulations suggest that employing a larger $L_a$ (i.e., at the cost of reducing $L_d$) benefits the system more. 
This is because there are more degrees of freedom in the analog precoding stage than that in the digital precoding stage. Hence, it is more beneficial to allocate more resources to designing the analog precoder.
In the simulations, we search for the optimal choice of $L_a$, starting from $L_a = L - 1$ and $L_d = 1$.} 
\section{Hybrid Precoding Design for Frequency-Selective OFDM Systems}
\label{sec:OFDM}
\begin{figure*}[!t]
        \centering
        \includegraphics[width=0.7\textwidth]{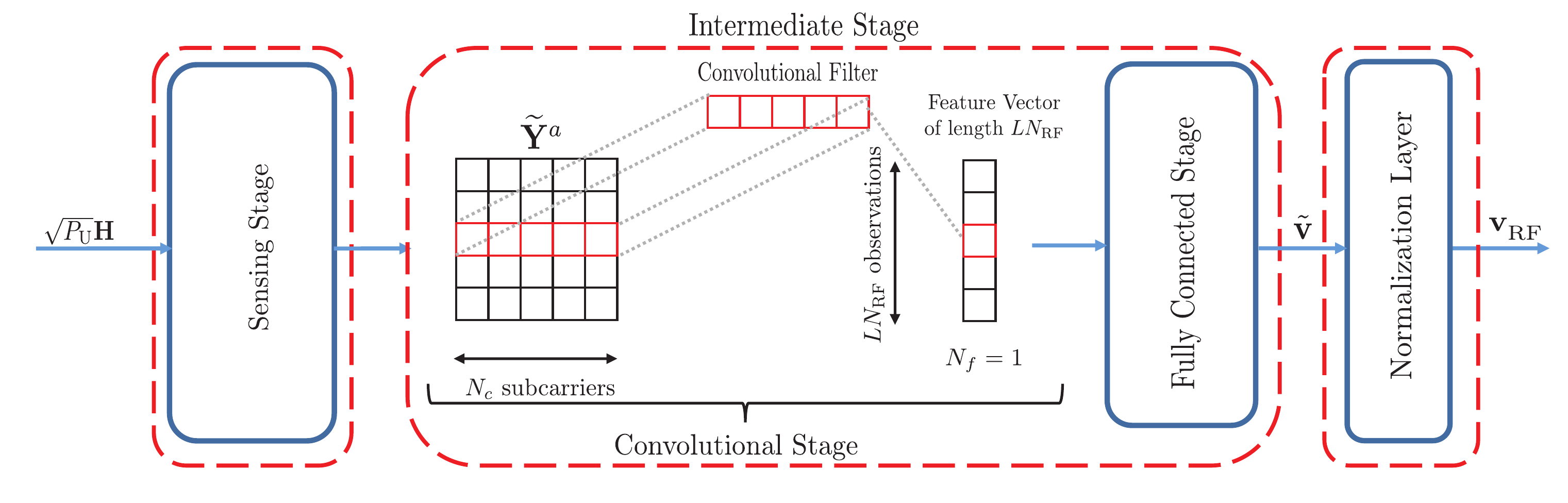}
        \caption{The proposed \editll{SU-DNN} for uplink sensing and downlink analog precoding design in multicarrier systems. \editr{For simplicity, the operation of the convolution stage is shown when $N_\text{f} = 1$.}}
   \vspace{-10pt}
        \label{fig:DNN_MC}
\end{figure*}
\editr{W}e now turn the attention to the more general OFDM based \editbbb{m}assive MIMO system. The extension of the proposed algorithm to \editbbb{the multicarrier case} should be made with \editbbb{the following considerations:} (i) \editll{A} common analog precoder should be designed 
\editbbb{for all channels over $N_c$ subcarriers,} and (ii) \editll{D}igital precoding should be performed on a per-subcarrier basis.

As before we divide the overall \editrev{precoding design} 
into an analog \editrev{precoding design}, in which the analog precoder is determined using a direct \editbbb{mapping} from the baseband \editrev{signal received in the first $L_a$ pilot transmissions}, and a digital \editrev{precoding design} that entails estimating the low-dimensional equivalent channel followed by linear precoding \editbbb{design}. To ensure that a common analog beamformer is \editbbb{appropriately} designed for the channels over the entire band, the \editll{SU-DNN} \editll{is} trained to learn a mapping from the baseband received pilots matrix over the entire frequency band (i.e., $\editbbb{\widetilde{\mathbf{Y}}_k}$ in~\eqref{eq:y_OFDM} for the $k$-th user) into the corresponding analog precoding vector. \editbbb{Further,} since digital precoding takes place on a per-subcarrier basis, 
the low-dimensional channel for each subcarrier is estimated and the corresponding digital precoder is subsequently determined using linear precoding. The details of the proposed scheme are provided in Algorithm 1.

\begin{algorithm} \small
\SetAlgoLined
\label{alg:OFDM}
\textbf{Input}: 
Number of pilot time frames $L_a$, $L_d$, uplink and downlink powers $P_\text{U}, P_\text{D}$, noise variance $\sigma^2$, 
\editbl{pretrained sensing matrices in the analog precoding design phase $\{\editr{\mathbf{W}}_\text{RF}^{(\ell)} \}_{\ell=1}^{L_a}$, and pretrained \editll{SU-DNN} $ \mathcal{G}_{\text{OFDM}}\left(\cdot; \mathbf{\Theta}_\text{OFDM}\right)$.} \\  
\textbf{Output}: The analog precoder $\mathbf{V}_{\text{RF}}$ and the digital precoders $\mathbf{V}_{\text{D}}[1], \ldots, \mathbf{V}_{\text{D}}[N_c]$. \\  
\textbf{Analog \editbbb{Precoding} Design}: \\    
 \For{$\ell = 1 \ldots, L_a$}{
   \editfi{-} BS \editbbb{sets} the analog \editbbb{sensing} to $\mathbf{W}_\text{RF}^{(\ell)}$\;
\editfi{-} Users send pilots $\mathbf{X}[1] , \ldots, \mathbf{X}[N_c]$ across subcarriers over $K$ OFDM symbols\;
\editfi{-} BS receives
      $\widetilde{\mathbf{Y}}^{(\ell)}_k = \sqrt{P_\text{U}}\mathbf{W}_{\text{RF}}^{(\ell)} \mathbf{H}_k + \mathbf{Z}^{(\ell)}_k$\;
}
  
  \For{$k = 1 \ldots, K$}{
 \editfi{-} BS \editbbb{constructs} $\widetilde{\mathbf{Y}}_k = \left[ \left(\widetilde{\mathbf{Y}}^{(1)}_k\right)^T, \ldots,  \left(\widetilde{\mathbf{Y}}^{(L_a)}_k\right)^T \right]^T$\;
 \editfi{-} BS computes $\mathbf{v}^{(k)}_\text{RF} = \mathcal{G}_\text{OFDM}\left(\widetilde{\mathbf{Y}}_k; \mathbf{\Theta}_\text{OFDM}\right)$\;
  }
 \editfi{-} BS determines $\mathbf{V}_\text{RF} = \left[ \mathbf{v}^{(1)}_\text{RF}, \ldots,  \mathbf{v}^{(K)}_\text{RF} \right]$\; 
\textbf{Digital \editbbb{Precoding} Design}: \\
    \editfi{-} BS \editbbb{sets} the analog \editbbb{sensing matrix} to  $\mathbf{V}_\text{RF}^H$\;
    
   \For{$\ell = L_a + 1 \ldots, L$}{
 \editfi{-} Users send pilots $\mathbf{X}[1] , \ldots, \mathbf{X}[N_c]$ across subcarriers  over $K$ OFDM symbols\;
 \editfi{-} BS receives
     $\widetilde{\mathbf{Y}}^{(\ell)}_k = \left[\tilde{\mathbf{y}}^{(\ell)}_k[1], \ldots, \tilde{\mathbf{y}}^{(\ell)}_k[N_c]  \right] = \sqrt{P_\text{U}} \mathbf{V}_\text{RF}^H \mathbf{H}_k + \mathbf{Z}^{(\ell)}_k$\;
}
  \For{$j = 1 \ldots, N_c$}{
 \editfi{-} BS \editbl{forms} $\widetilde{\mathbf{\Upsilon}}^{(\ell)}[j] \triangleq \left[ \tilde{\mathbf{y}}^{(\ell)}_1[j], \ldots,  \tilde{\mathbf{y}}^{(\ell)}_K[j] \right], \forall \ell$\; 
 \editfi{-} BS determines the LMMSE estimate
   $\hat{\mathbf{H}}_{\text{eq}}[j] =  \frac{\sqrt{P_\text{U}}}{P_\text{U}L_d + \sigma^2} 
     \sum_{\ell = L_a + 1}^{L} \widetilde{\mathbf{\Upsilon}}^{(\ell)}[j]$\;
     \editfi{-} BS determines  $\mathbf{V}_\text{D}[j]$ using ZF or WMMSE on $\hat{\mathbf{H}}_{\text{eq}}[j]$.
  }
 \caption{Hybrid Precoding for OFDM Systems}
\end{algorithm}

\editf{Analogous to the SU-DNN \edit{for} the single-carrier case, the network architecture in the OFDM case consists of a sensing stage, an intermediate stage and a normalization layer. The main difference is that the DNN in the OFDM case is designed to map the received pilots across all frequencies into a single analog precoder. To accomplish this, we propose an intermediate stage consisting of a convolutional layer followed by an FC stage. We utilize a convolutional layer to reduce the input dimension and to exploit the correlation of the channel across the subcarriers, thereby reducing the computational complexity of the subsequent FC stage. In particular, we apply $N_f < N_c$ convolutional filters to the columns of the recieved pilot matrix $\tilde{\mathbf{Y}}_k$, where each convolutional filter is a 1-D horizontal stripe with length $N_c$ and trainable parameters. Each filter output is then followed by a ReLU activation. 
This produces a convolution output consisting of $N_f$ feature vectors each of length $LN_\text{RF}$. The feature vectors are then mapped into the unnormalized precoding vector using the subsequent FC stage. The overall DNN architecture is illustrated in \figurename~\ref{fig:DNN_MC}, where the sensing stage and normalization layer are analogous to their counterparts for the DNN of single-carrier systems.}
Finally, we \editll{use} the following loss function:
\begin{equation}\small
\label{eq:loss_function_OFDM}
\mathcal{L}(\mathbf{V}_{\text{RF}}) =  -  \sum_{k=1}^K \sum_{j=1}^{N_c} \log_2\left(1 + \frac{P_\text{D}}{M K \sigma^2} \left\lvert \bh_k^H[j] \mathbf{v}_{\text{RF}}^{(k)} \right\rvert^2 \right),
\end{equation}
which generalizes the loss function~\eqref{eq:loss_function}.
\section{Numerical Results}
\label{sec:numerical}
In this section, we evaluate the performance of the proposed hybrid precoding scheme against existing benchmarks and investigate its ability to generalize in various system parameters. 
\subsection{Parameter Configuration and Implementation Details}
We consider a massive MIMO \editb{system} in TDD where the BS is equipped with $N_\text{RF} = 4$ RF chains and a two-dimensional UPA with 
$M_h = M_v = 8$. Thus, the total number of BS antennas is $M = 64$. The antenna separation in both directions is set to $\lambda/2$. For OFDM systems, the user channels follow the model 
introduced in Section~\ref{sec:chl_mdl} with $L_p = 4$ \editbbb{paths,} a maximum delay spread $d_\text{max} = 4$\editr{,} and $N_c = 128$ subcarriers. \editrev{The pilot matrices for different subcarriers have the form $\mathbf{X}[j] = e^{\imath \phi[j]}\mathbf{X}$ with a randomly chosen rotation phase $\phi[j]$ and $\mathbf{X}$  set to be a  $K \times K$ DFT matrix. 
The same $\phi[j]$'s are used across all the time frames in the entire pilot phase.} 
The channel paths are assumed to be independent and identically distributed (i.i.d.) with complex Gaussian path gains $\alpha_{\ell, k} \sim \mathcal{CN}(0, 1)$ and uniform path delays $\tau_{\ell, k}$ over the interval $[0, d_\text{max} T_s]$, where $T_s = \tfrac{1}{1760} \mu s$ \cite{Rodriguez2018}. The azimuth and elevation AoDs follow a uniform distribution over the interval $[-\pi/2, \pi/2]$ and the pulse shaping filter is raised cosine with a roll-off factor of~$0.8$. For the single-carrier system, we adopt the frequency-flat mmWave model introduced in Section~\ref{sec:sc} by setting $d_\text{max} = 0$ and $N_c = 1$. 
Finally, we define the uplink SNR as $\text{SNR}_\text{UL} = 10 \log{\tfrac{P_\text{U}}{\sigma^2}}$ and the downlink SNR as $\text{SNR}_\text{DL} = 10 \log{\tfrac{P_\text{D}}{\sigma^2}}$. 


We implement the proposed DNNs 
using TensorFlow~\cite{tensorflow2016}. Since \editr{most deep learning} libraries do not support complex operations, \editr{we represent all complex quantities using their real representations} 
and implement complex multiplications 
using real addition and multiplication. We set the number of FC intermediate layers of the \editr{single-carrier/OFDM SU-DNNs} to $R = 3$, with dense layers of widths 1024, 512, and 256, respectively. 
For faster convergence, each dense layer is preceded by a batch normalization layer. 
In addition, for the OFDM neural network, we set the number of convolutional filters $N_f = 16$\editbbb{, which} leads to a dimensionality reduction of the input space by a factor of $8$ for a system with $N_c = 128$. 
We train the models using Adam optimizer~\cite{adam2014} with minibatches of size \editrev{$B = 500$} over \editf{many} \editt{($\sim1000$)} epochs. The learning rate \editbbb{is} initialized at $10^{-3}$ and progressively decreased by a factor of $2$ every $100$ epochs. 
\editrev{The size of the training set is set to be $50{,}000$ channel realizations, where each realization corresponds to the channels of $K = 4$ users. Hence, for the purpose of training a single SU-DNN, the overall training set size equals $200{,}000$.} We use a validation set of size $1000$ to monitor the performance and keep the model parameters that achieved the best generalization. \editrev{Finally, the test set used in the simulations consists of $10{,}000$ channel examples.} 
\begin{table*}[t]
\renewcommand{\arraystretch}{1.3}
\captionsetup{justification=centerlast, labelsep=newline, font=sc,
textfont=footnotesize, labelfont=small}
\caption{Relative frequency of the event that the sum rate of the proposed scheme exceeds that of~\cite{ma2020}.}
\label{tab:table-name}
\centering
\begin{tabular}{ |l||c|c|c|c|c| }
 \hline
 Pilot Length ($L$) & $L = 2$ & $L = 4$ & $L = 6$ & $L = 8$ & $L = 10$ \\ 
 \hline
 Analog Pilot Phase ($L_a$) & $L_a = 1$ & $L_a = 3$ & $L_a = 4$ & $L_a = 6$ & $L_a = 7$ \\
 \hline
 $\text{Pr}\left(\sum R_\text{Proposed} > \sum R_\text{DLCS} \right)$ & $ 69.6\%$ & $78.2\%$ & $91.5\%$ & $91.7\%$ & $91\%$ \\
 \hline
\end{tabular}
\end{table*}

\begin{figure*}[t]
        \centering
        \begin{subfigure}[t]{0.32\textwidth}
        \centering
        \includegraphics[width=\columnwidth]{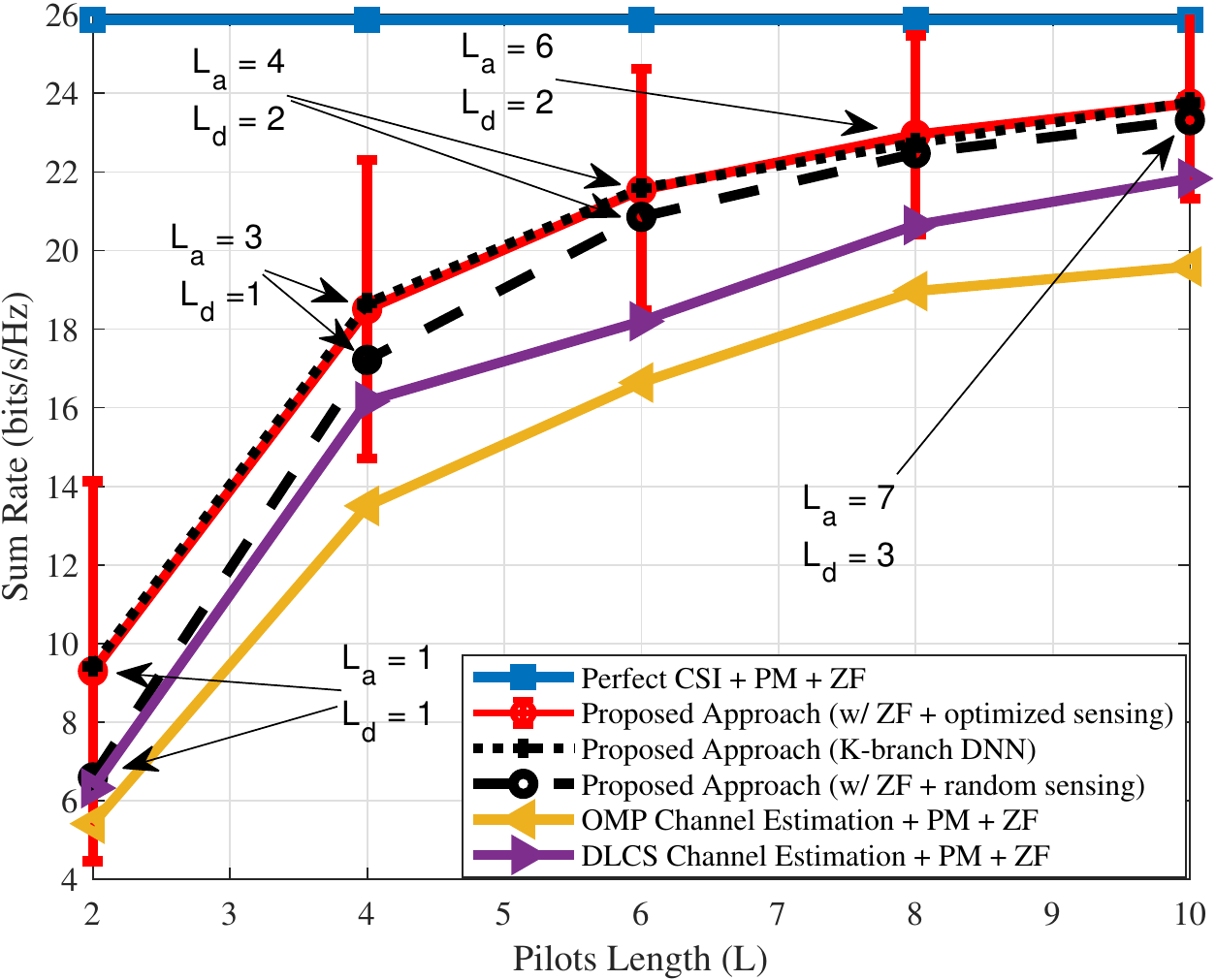}
        \caption{Sum rate vs pilot length for a single-carrier system with $ \text{SNR}_{\text{DL}} = 10~\text{dB}$.} 
        \label{fig:compare_1}
        \end{subfigure}
        \hfill
        \begin{subfigure}[t]{0.32\textwidth}
        \centering
        \includegraphics[width=\columnwidth]{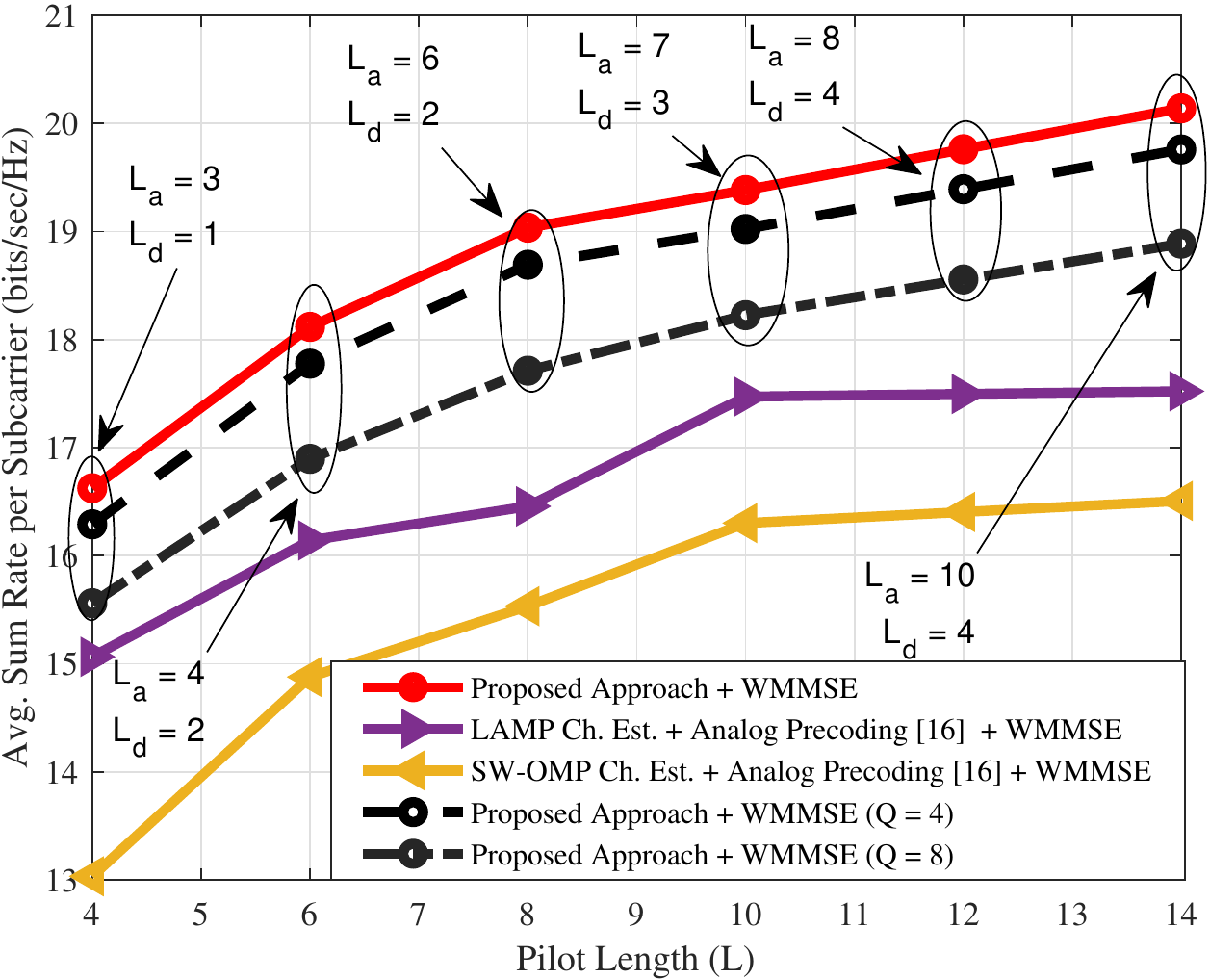}
        \caption{Sum rate vs pilot length for a multicarrier system with $N_c = 128$ subcarriers.}
        \label{fig:compare_ofdm}
        \end{subfigure}
        \hfill
        \begin{subfigure}[t]{0.32\textwidth}
        \centering
        \includegraphics[width=\columnwidth]{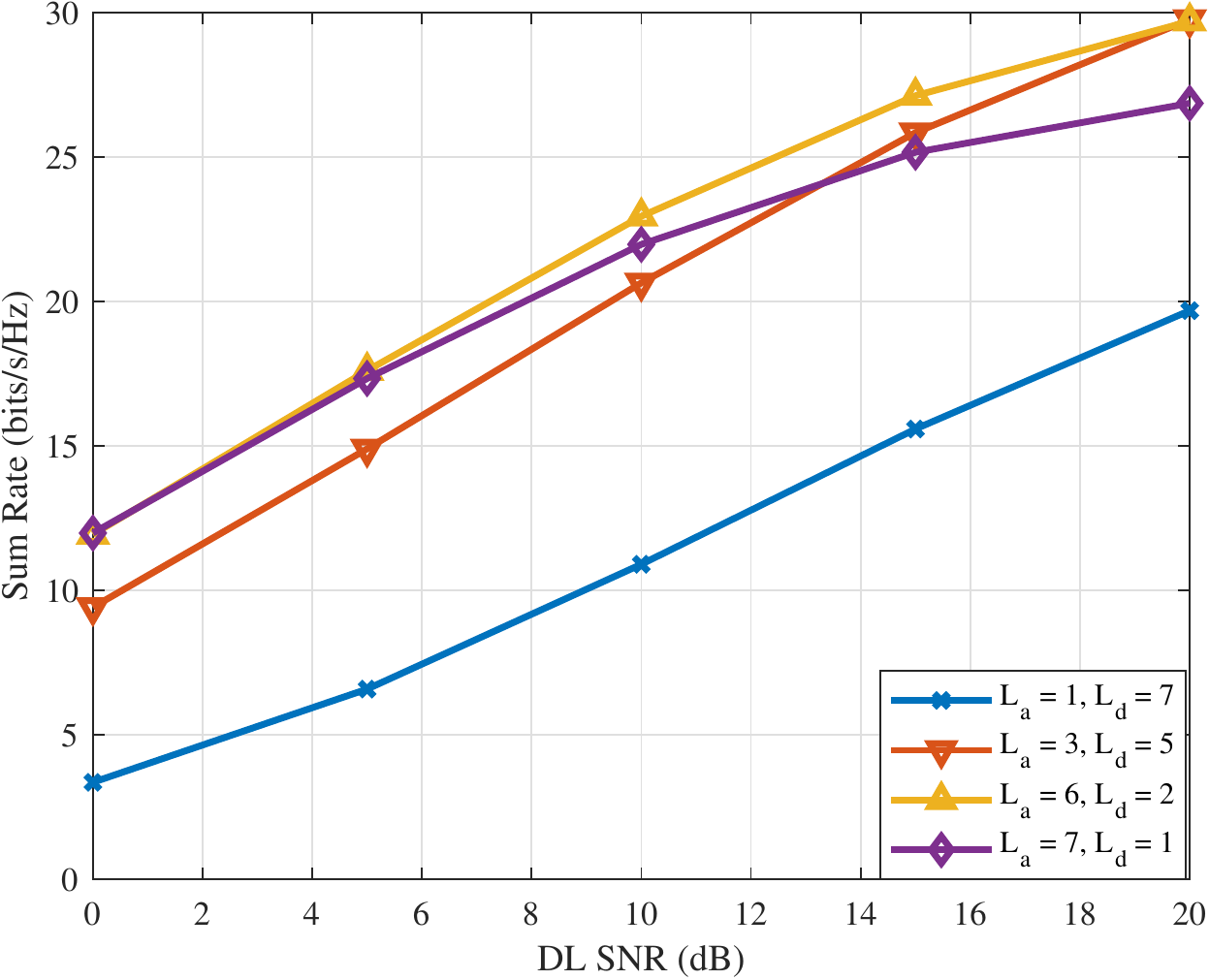}
        \caption{Effect of pilot allocation in a single-carrier system with $L = 8$.}
        \end{subfigure}
        
        \caption{Performance analysis 
        of the proposed approach 
        when $K = N_\text{RF} = 4$ and $\text{SNR}_\text{UL} = 10$dB.} 
        \label{fig:compare}
\end{figure*}
\subsection{\editrev{Sum-Rate Performance Analysis}}
\label{sec:perform_comp}
We first analyze the performance of the proposed hybrid precoding scheme which bypasses the channel estimation step for the analog precoding design in a multiuser setup with $K = 4$ users and $\text{SNR}_\text{UL} = \text{SNR}_\text{DL}= 10$dB. In \figurename~\ref{fig:compare}(a), we plot the sum rate against the number of pilot frames $L$ assuming a single-carrier mmWave massive MIMO setup.  For the proposed approach, we indicate the values of $L_a$ and $L_d$ \editbbb{on the figure}. As benchmarks, we consider downlink precoding using \editf{phase matching}~\cite{liang2014low} for analog precoding and ZF for digital precoding using perfect/imperfect CSI. 
For the channel estimation schemes in the imperfect CSI case we use either the OMP \editbl{algorithm}~\cite{Lee2016} or \editr{the 
 deep learning compressed sensing (DLCS) approach~\cite{ma2020}.} 
 \editbbb{It can be seen in \figurename~\ref{fig:compare}}(a) that the proposed scheme significantly outperforms channel \editbbb{ recovery} based counterparts. For example, the proposed approach \editbbb{with $L = 8$} achieves over $85\%$ of the total sum rate of the full CSI systems, while the DLCS does so with $L = 10$. 
\editbbb{This} indicates over $20\%$ saving in pilot overhead relative to the conventional channel recovery based schemes. This supports the main claim that the performance of the hybrid precoding system can be enhanced by bypassing explicit channel estimation for the analog precoding design. \editrev{In addition, we show the variations around the mean for the proposed scheme using error bars that represent one standard deviation from the average rate. 
Finally, we also plot the performance of the proposed scheme when: (i) the sensing
$\mathbf{W}_\text{RF}^{a}$ is random and (ii) each user uses a different DNN with different parameters~$\mathbf{\Theta}_k$. From these comparisons, it is observed that the effect of optimizing $\mathbf{W}_\text{RF}^{a}$ is noticeable only at a very short pilot length. Further, tying the DNN parameters does not result in performance loss in the i.i.d. channels case.}

\editrev{To examine whether the proposed scheme is always better than the channel estimation schemes for all channel realizations, in Table~\ref{tab:table-name}, we report the relative frequency of the event that the sum rate of the proposed scheme exceeds the sum rate of the DLCS scheme for different values of $L$. From the table, it is seen  that the proposed scheme is superior a majority of the time. 
}

Next, we consider the sum-rate comparison for the multicarrier setup. This comparison is shown in~\figurename~\ref{fig:compare}(b). 
For the baselines, we \editll{perform} channel estimation according to SW-OMP~\cite{Rodriguez2018} and the data-driven LAMP network~\cite{MALAMP2021}\editll{,} and utilize the covariance averaging scheme of \cite{SohrabiOFDM} for analog precoding and WMMSE for digital precoding. \editbbb{From~\figurename~\ref{fig:compare}(b)}, we \editbbb{again} observe that the performance of the proposed data-driven scheme is \editbbb{superior to the} channel recovery based ones. Further, we also plot the sum rate of the proposed precoding approach assuming finite-resolution phase shifters. We observe that the performance of \editbbb{the proposed approach with only} 2-bit phase shifters 
can already exceed the performance of the channel recovery based approaches with infinite-resolution phase shifters. Moreover, the performance of \editbbb{the proposed \editll{method} with 3-bit} phase shifters \editll{already approaches} that of the infinite-resolution counterpart. 

\editrev{Finally, we examine the performance of the proposed scheme under different pilot allocation settings. In~\figurename~\ref{fig:compare}(c), we set $L = 8$ and plot the sum rate against the downlink SNR for the cases of $L_a \in \{1, 3, 6, 7\}$, where it can be seen that the case of $L_a = 6$ corresponds to the best performance. The cases $L_a \in \{4, 5\}$ exhibit roughly the same performance as the $L_a = 6$ case, and are omitted from the figure for clarity. In this example, it can be observed that the proposed scheme incurs a performance loss whenever the pilots are not allocated optimally. Moreover, it is observed that allocating more pilots to the analog precoding design yields better performance.}
\subsection{Generalizability Results}
\label{gen_res}
In this section, we seek to examine how well the system \editbbb{performs} under parameter settings other than the ones used for training. This is \editbbb{crucial} \editt{because} \editbbb{wireless communication systems} are typically far from static and their parameters \editbbb{constantly change over time}. The main goal of this section is to show that the proposed data-driven scheme \editbbb{can} maintain a \editbbb{robust operation against \editll{variations} in system parameters.} 
 
\begin{figure*}
        \centering
        \begin{subfigure}[t]{0.32\textwidth}
        \includegraphics[width=\textwidth]{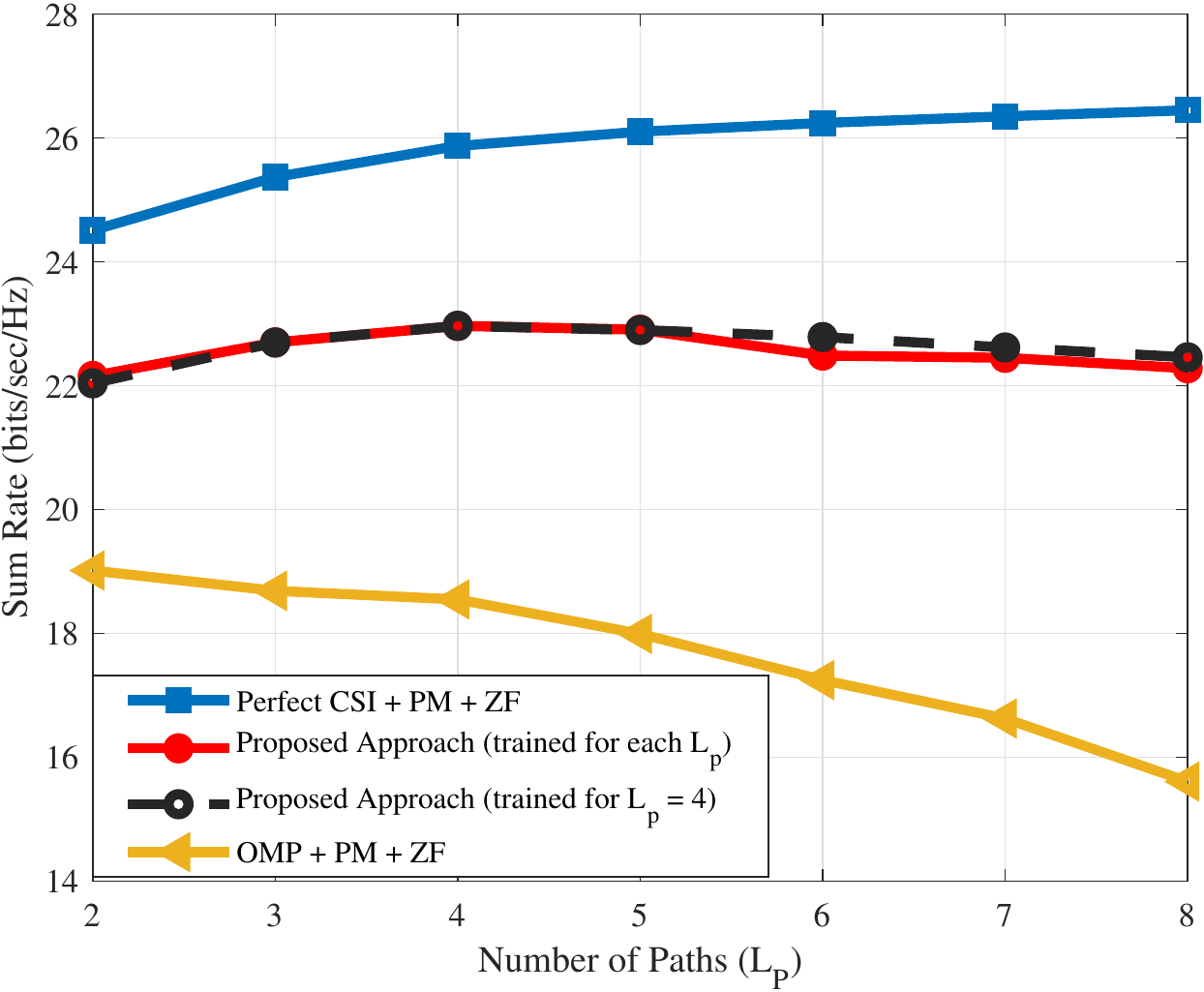}
        \caption{Generalizability in the number of paths. We fix $\text{SNR}_\text{UL} = 10~\text{dB}$, and vary $L_p$.}
        \label{fig:geninLp}
        \end{subfigure}
        \hfill
        \begin{subfigure}[t]{0.32\textwidth}
        \includegraphics[width=\textwidth]{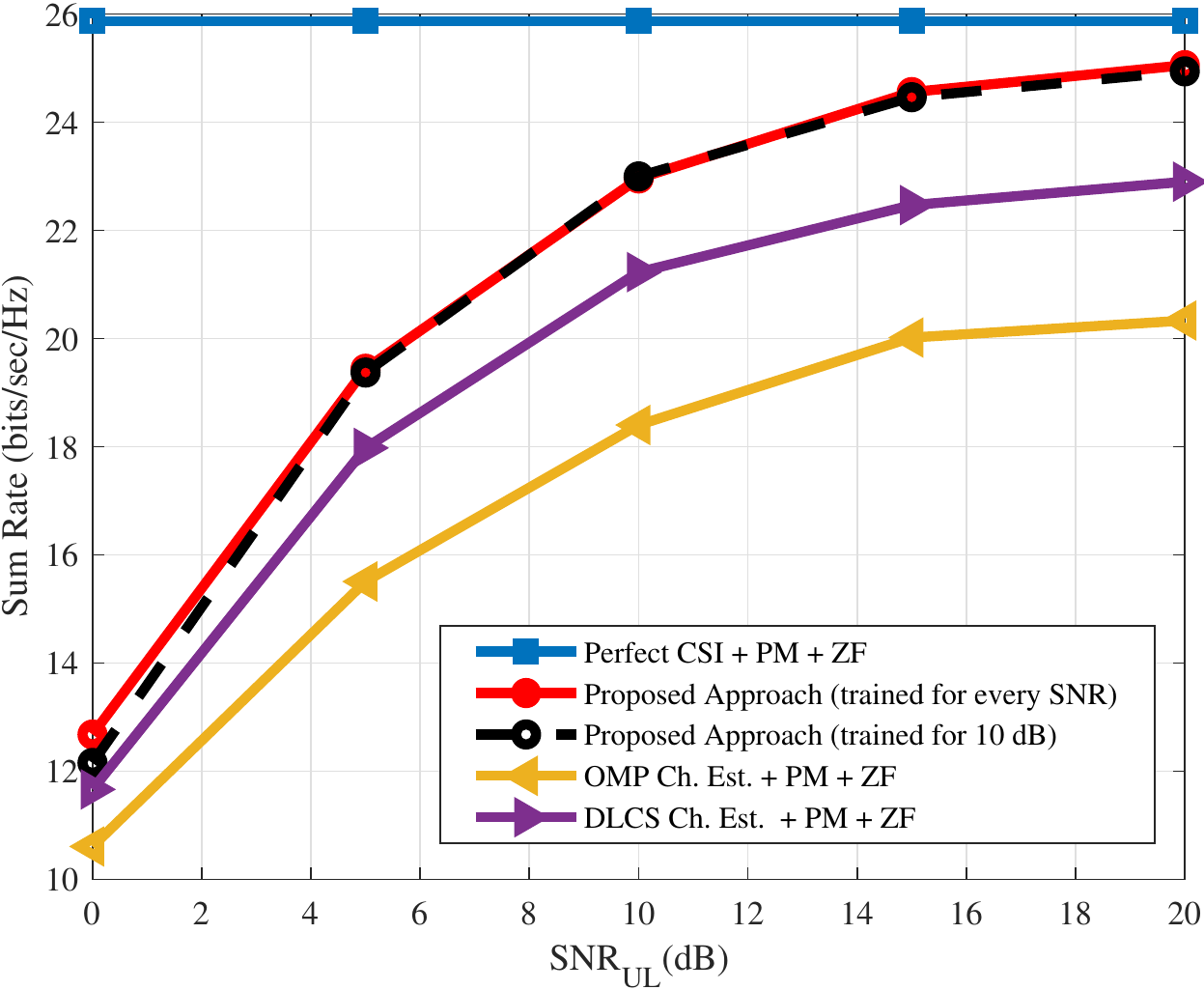}
        \caption{ Generalizability in the uplink SNR. We fix $L_p = 4$, and vary $\text{SNR}_{\text{UL}}$.}
        \label{fig:geninULSNR}
        \end{subfigure}
        \hfill
        \begin{subfigure}[t]{0.32\textwidth}
    \includegraphics[width=\textwidth]{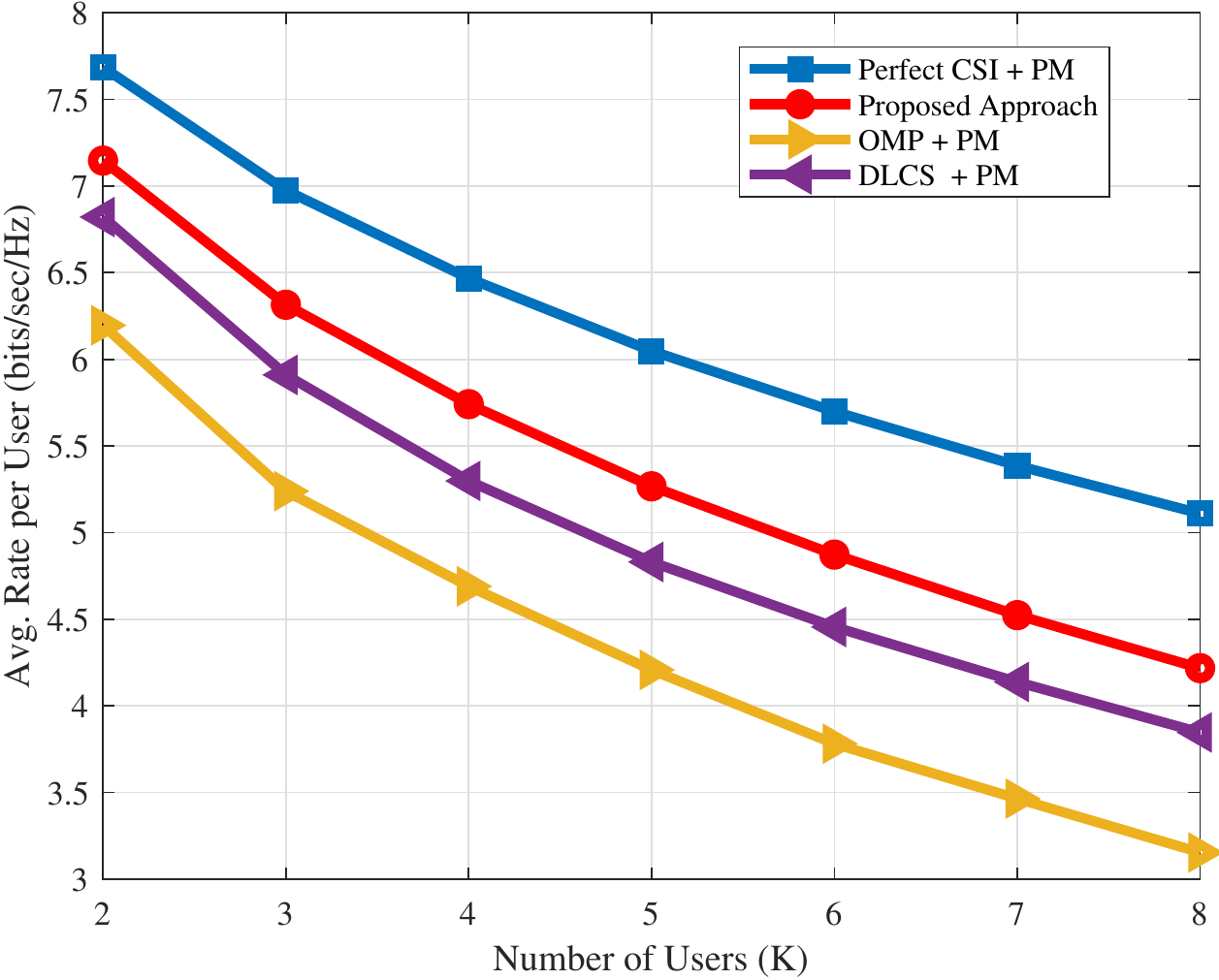}
    \caption{Generalizability in the number of users for fixed $L_p$ and $\text{SNR}_\text{UL}$.}
    \label{fig:compare_K}
    \end{subfigure}
    
        \caption{Generalizability of the single-carrier DNN in different parameters. We set $N_{\text{RF}} = 4$, and $\text{SNR}_{\text{DL}} = 10 $ dB.}
        \label{fig:SC}
\end{figure*}

\begin{figure*}
\begin{subfigure}[t]{0.32\textwidth}
        \includegraphics[width=\textwidth]{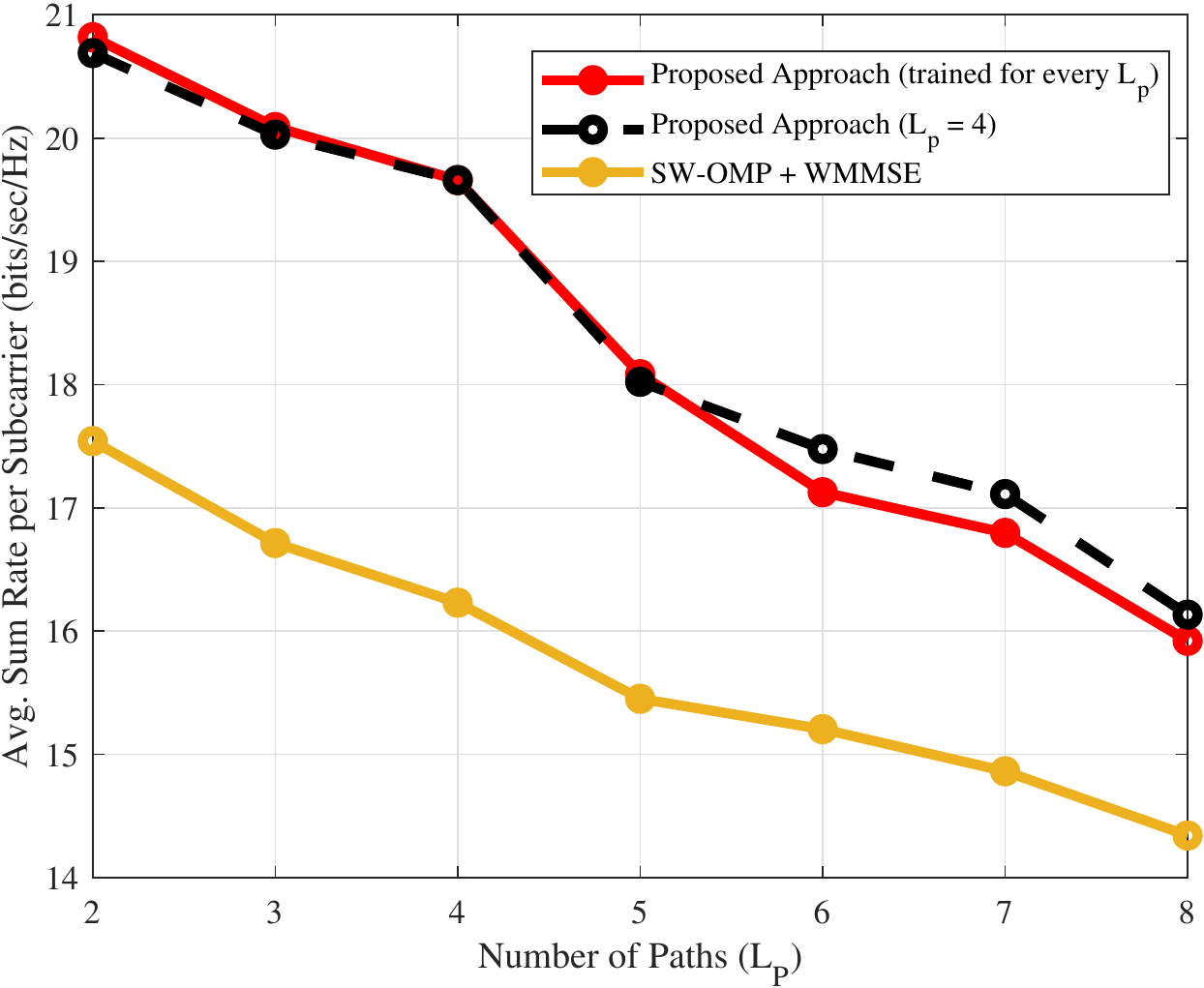}
        \caption{Generalizability in the number of paths. We fix $\text{SNR}_\text{UL} = 10~\text{dB}$, and vary $L_p$.}
        \label{fig:geninLp_OFDM}
        \end{subfigure}
        \hfill
         \begin{subfigure}[t]{0.32\textwidth}
        \includegraphics[width=\textwidth]{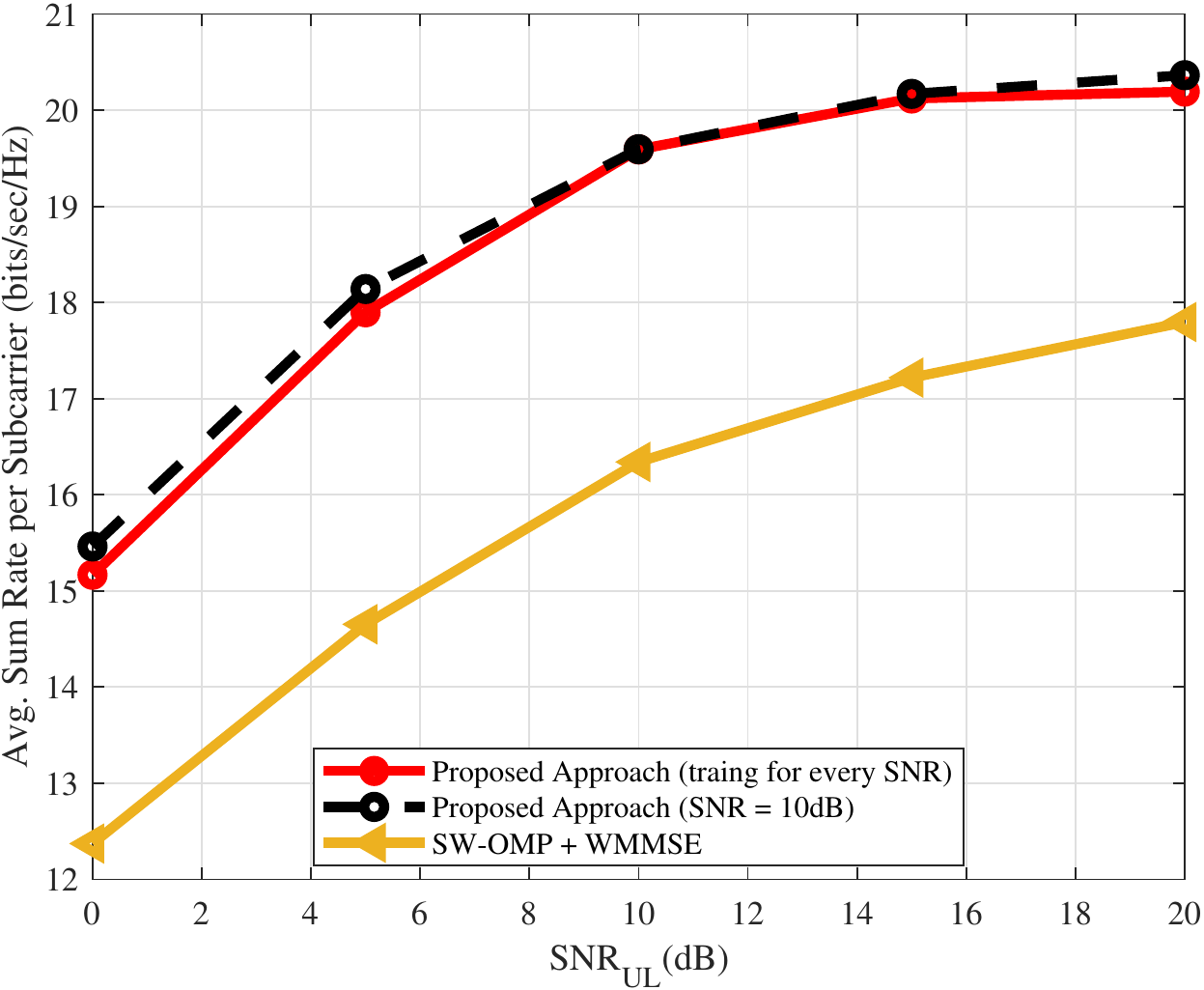}
        \caption{Generalizability in the uplink SNR. We fix $L_p = 4$, and vary $\text{SNR}_{\text{UL}}$.}
        \label{fig:geninULSNR_OFDM}
        \end{subfigure}
        \hfill
            \begin{subfigure}[t]{0.32\textwidth}
    \includegraphics[width=\textwidth]{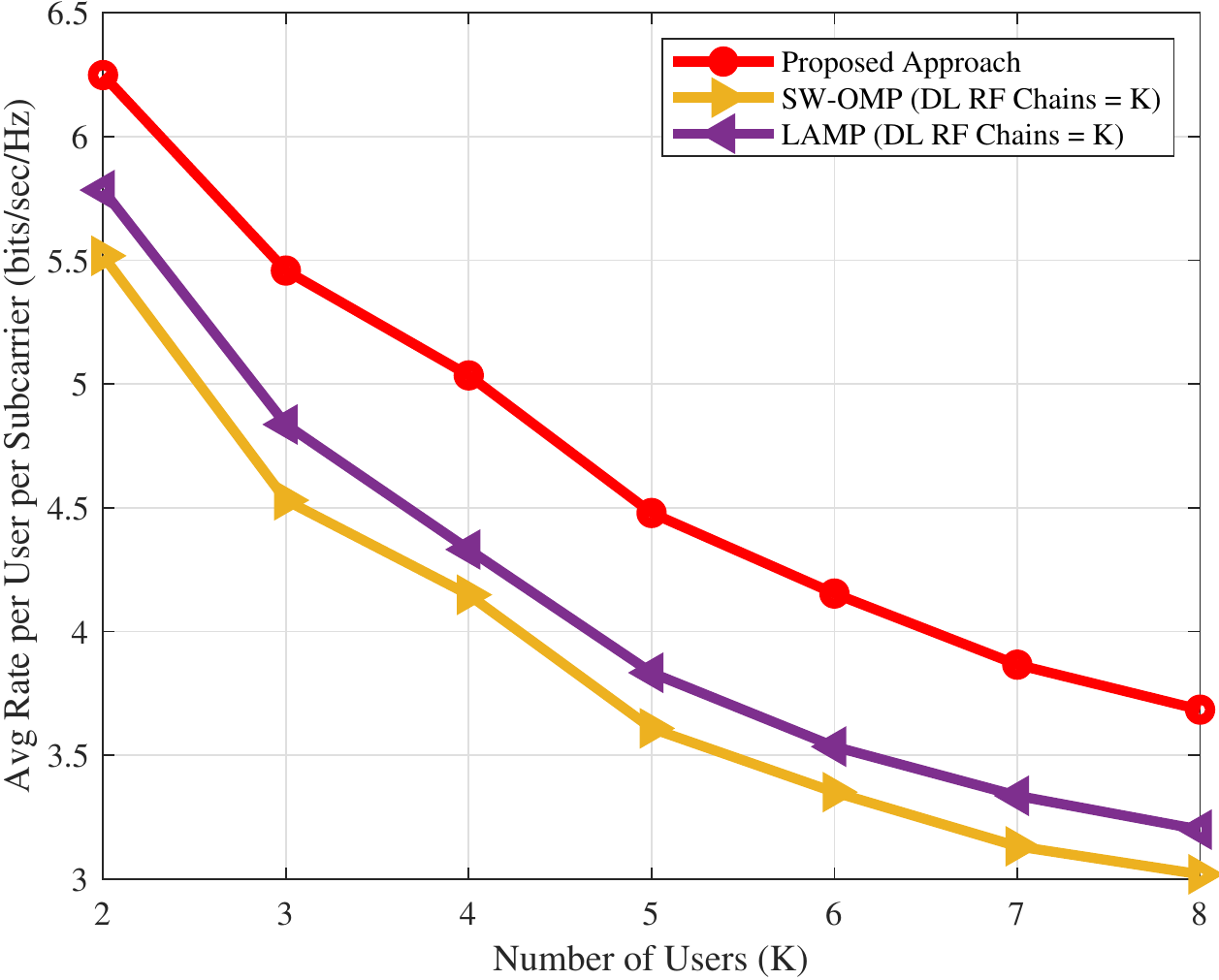}
    \caption{Generalizability in the number of users for fixed $L_p$ and $\text{SNR}_\text{UL}$.}
    \label{fig:compare_K_multi}
    \end{subfigure}
        
        \caption{Generalizability of the multicarrier DNN in different parameters. We set $N_{\text{RF}} = 4$, $N_c = 128$, and $\text{SNR}_{\text{DL}} = 10$ dB.}
        \label{fig:MC}
        \vspace{-10pt}
\end{figure*}

First, we investigate the \editbbb{generalizability} of the proposed scheme in the number of paths. To this end, we consider two scenarios for the proposed approach. In the first scenario, the DNN is trained and tested using \editbbb{datasets} generated from the same distribution. This experiment is repeated 
\editbbb{for different values of the number of paths, i.e.,} $L_p \in \{2, \ldots, 8\}$. Hence, this scenario represents an \editrev{unrealistic baseline} where \editrev{a different DNN is trained for different values of $L_p$.} 
In the second scenario, \editrev{which represents the actual performance of the proposed approach}, we train the DNN using a training set with $L_p = 4$ and evaluate the performance on the test sets with \editbbb{$L_p \in \{2, \ldots, 8\}$.} 
We set $L_a = 6$ and $L_d = 2$ for the single-carrier setup and $L_a = 8$ and $L_d = 4$ for the multicarrier setup. All other parameters remain the same as those given in Section~\ref{sec:perform_comp}. \editll{W}e plot the sum rate against the number of paths in \figurename~\ref{fig:SC}(a) for the single-carrier case and in \figurename~\ref{fig:MC}(a) for the multicarrier \editbbb{case}. 
For comparison purposes, we also include the performance of the perfect CSI case and the imperfect CSI case with OMP for channel estimation in the single-carrier system, and SW-OMP in the multicarrier system. It can be seen that there is no tangible loss in \editrev{the performance of the proposed approach} 
relative to the \editrev{baseline}
. This indicates that the proposed DNN is able to maintain a robust operation against \editll{variations} in $L_p$.

Next, we numerically study the \editb{degradation} in performance due to mismatch in the uplink SNR. \editbbb{Analogous to the previous simulation,} we consider two scenarios \editbbb{in this experiment.} 
Specifically, in the first scenario\editrev{, representing the baseline, the DNN is retrained as}  
the uplink SNR is varied \editrev{between} 
$0-20$dB. In the second scenario, \editrev{representing the actual performance of the proposed scheme}, the DNN is trained using a training set generated for a fixed uplink SNR of $10$dB. The results are shown in \figurename~\ref{fig:SC}(b) for the single-carrier system and in \figurename~\ref{fig:MC}(b) for the multicarrier system. 
Once again, we see that there is only a negligible loss in \editrev{the performance of the proposed scheme relative to the baseline}
, thereby indicating that the proposed design is able to perform well even in the presence of variations in the uplink SNR. 

Finally, we demonstrate the applicability of the proposed approach to systems serving different number 
of users. 
First, we study the case where all the users share the same channel distribution, the more general case in which the users do not share the same channel distribution is discussed in the next \editbl{subsection}. \editbbb{In the former case, we train one single-user neural network on the channel examples of all $K$ users for the proposed scheme. 
After training, this \editll{SU-DNN} is then duplicated across the $K$ branches of the overall DNN.} 
\editrr{We examine the single-carrier case in \figurename~\ref{fig:SC}(c) with $L = 3$, $L_a = 2$ and $L_d = 1$, and the multicarrier case in \figurename~\ref{fig:MC}(c) with $L = 7$, \editrev{$L_a = 5$ and $L_d = 2$.} In both simulations, we set the number of paths $L_p = 4$, and $\text{SNR}_{\text{UL}} = \text{SNR}_{\text{DL}} = 10$dB. } 
To ensure that the number of RF chains \editf{is always equal to} the number of users, we select only $K$ RF chains for downlink precoding.  This assumption is needed for the proposed scheme as well as the PM design.
In both cases, we observe that the proposed approach provides 
better performance than the channel recovery based approaches\editt{, regardless of the number of users.} 

\subsection{\editt{System\editf{-}Level Performance}}
\editt{The simulation results presented so far pertains to the sum rate objective. To evaluate the system\editf{-}level performance of the proposed algorithm \editf{and to} account for fairness across the users, we }
\editrr{examine the performance of the proposed approach in an urban micro cell with $200$ meters radius and $2000$ potential users}
. The users are placed randomly in a circular region within distances between $30$ to $200$ meters from the BS. \editbl{The antenna gain and transmit power at the BS are $0$dBi and \editr{$40$dBm}. Similarly, the antenna gain and transmit power at the users are $15$dBi and \editr{$30$dBm}.}
Further, it is assumed that the users experience both large-scale fading and small-scale fading. The small-scale fading component follows the frequency-selective mmWave model presented in Section~\ref{sec:chl_mdl} with $L_p = 4$, whereas the large-scale fading component follows the floating intercept model derived in\cite{maccartney2013path} from $38$~GHz empirical measurements.
\editbl{The allocated bandwidth is $10$MHz and the \editr{noise spectral density is $-173.8$~dBm/Hz}}.

We assume that communication occurs in a time slotted fashion. In each time slot, $N_\text{RF} = K = 4$ users are scheduled randomly. Further, we consider weighted sum rate as the metric of performance, \editrev{i.e., $\sum_{k} w_k R_k$,}
where the weights \editrev{$w_1, \ldots, w_k$} are inversely proportional to the long term average of the scheduled users' rates in previous time slots. 
\editrev{That is, $w_k = \tfrac{1}{\bar{R}^t_k}$, where $\bar{R}^t_k$ is the average rate of user $k$ up to the $t$-th time slot.} For the proposed design, we choose $L_a = 10$ and $L_d = 4$, and set $L = 14$ for the other precoding schemes. Moreover, for the analog precoder design we train one \editll{SU-DNN} (using the loss function in~\eqref{eq:loss_function_OFDM}) on a training set sampled randomly from the channel \editf{distributions} of all users in the cell. After that, we evaluate the performance using $4$ identical copies of the trained DNN. Finally, we use WMMSE \editf{to design the digital precoder for both} the proposed and the baseline approaches.

\begin{figure}[!t]
    \centering
    \includegraphics[width=0.4\textwidth]{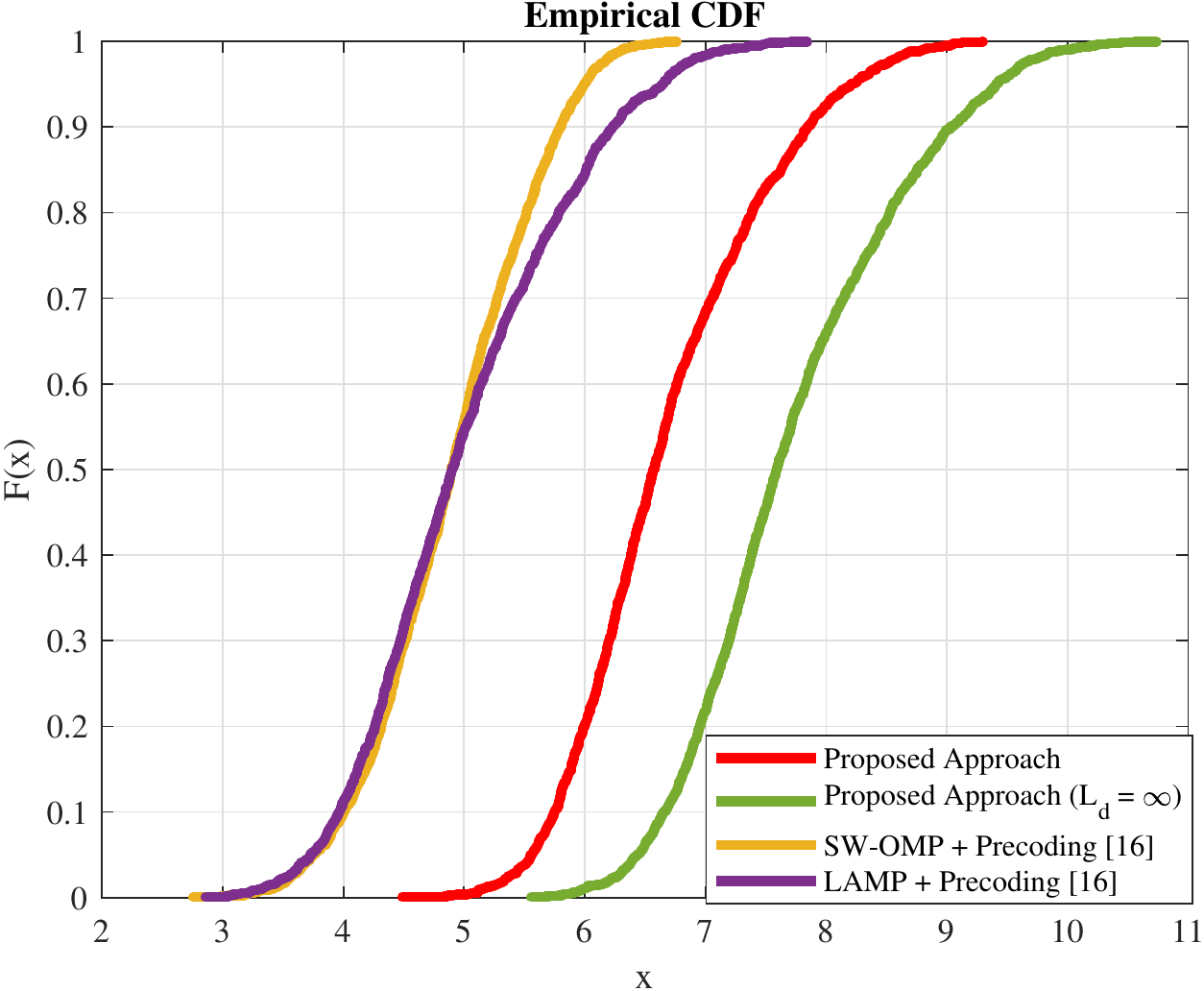}
    \caption{Empirical CDF of the average user rate under different hybrid precoding schemes in an urban cell scenario. \editll{We set $M = 64$, $N_\text{RF} = K = 4$, $L_a =10$, $L_d = 4$, and $L = 14$.}
    }
    \label{fig:cell}
    \vspace{-10pt}
\end{figure}

\figurename~\ref{fig:cell} shows the empirical cumulative distribution function (CDF) of the average rate per user for different hybrid precoding schemes. 
\editll{I}t can be observed that the average user rate under the proposed scheme is much higher than the average user rate under the channel recovery based precoding schemes. This suggests that the \editll{SU-DNN} with sufficient complexity can generally learn a good mapping for the analog precoding design despite being trained under different channel conditions. In other words, sharing the weights across the branches of the DNN architecture can still ensure generalizability in the number of users even when the users do not share the same channel distribution. 
\editll{Moreover,} the superior performance of the proposed scheme suggests its suitability for maximizing not only the sum rate but a more general network utility function. 
This fact can be seen by noting that the loss function in~\eqref{eq:loss_function} can be regarded as a universal measure of performance for the analog precoding design since it encourages the analog beamformer to match the channel phases\editf{, while allowing the subsequent digital precoder to alleviate the interuser interference}. 
Finally, comparing the CDF curves of the proposed approach when $L_a = 10$ and $L_d = \infty$ indicates that the proposed design can be further improved by employing longer $L_d$. 

\begin{figure}[!t]
    \centering
    \includegraphics[width=0.4\textwidth]{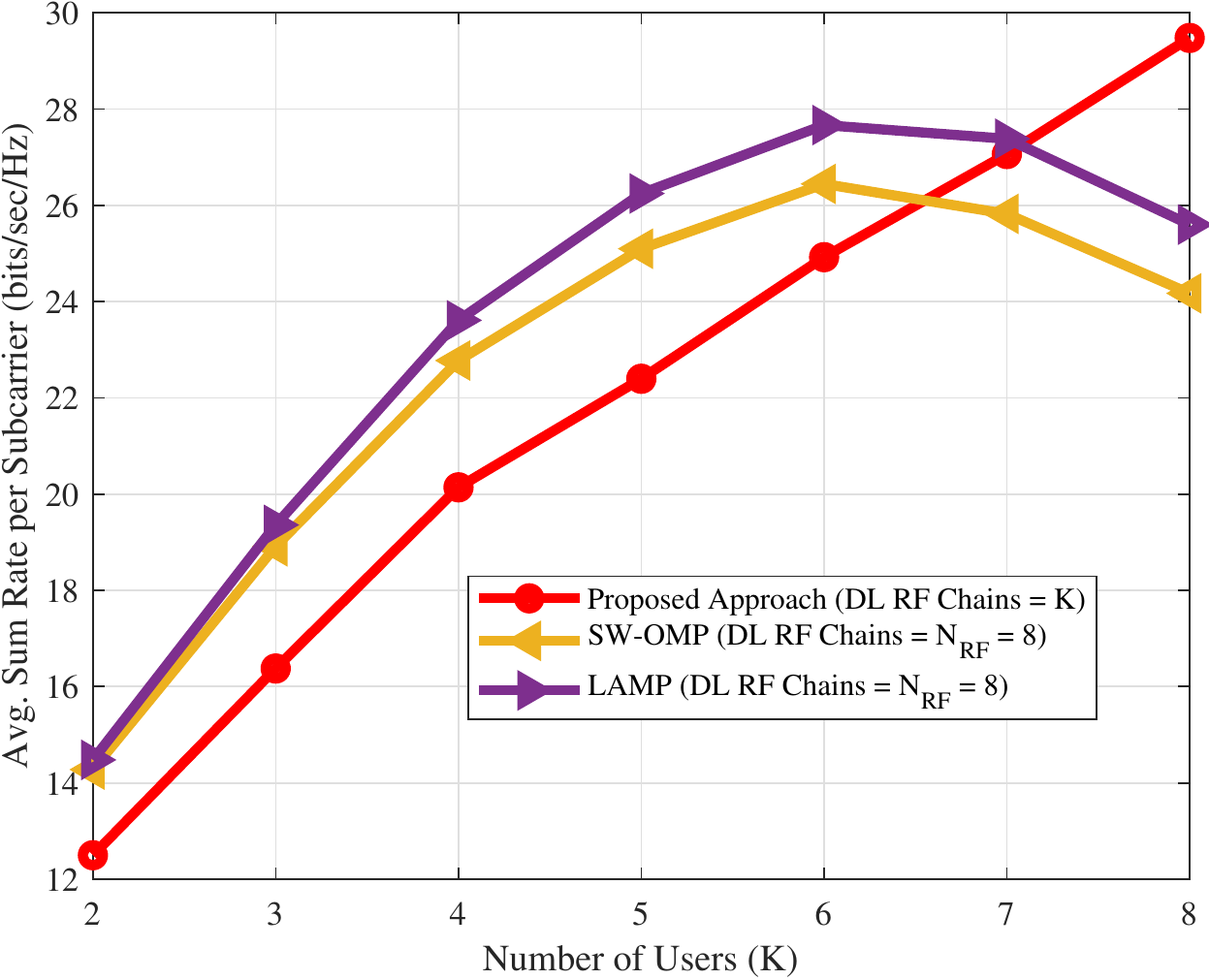}
    \caption{\editrr{Performance of the proposed scheme when $N_\text{RF} \geq K$. We set $M = 64$ and $L =7$, with $L_a = 5$ and $L_d = 2$. }
    }
    \label{fig:sumratevsK}
    \vspace{-10pt}
\end{figure}

\subsection{\editrr{Performance \editf{when}
$N_\text{RF} > K$}}
\editrr{As a final simulation, we investigate the performance of the proposed approach when the number of RF chains exceeds the number of users. To this end, \editf{we} fix $N_\text{RF} = 8$ and plot the performance for different values of $K \leq 8$. We remark here that because of our initial assumption that $N_\text{RF} = K$, we can only utilize $K$ out of $N_\text{RF}$ RF chains for downlink precoding using the proposed design. 
In contrast, the channel recovery based approaches can utilize all $N_\text{RF}$ RF chains for downlink precoding since the covariance averaging approach of~\cite{SohrabiOFDM} does not require this assumption. \figurename~\ref{fig:sumratevsK} shows the comparison for different precoding schemes. We observe that the proposed scheme can \editf{eventually} outperform the channel recovery based approaches \editf{when the system is fully loaded, but it achieves a lower sum rate when $K < N_\text{RF}$}. 
This is because the channel recovery based schemes can take advantage of the increased degrees of freedom (offered by utilizing all $N_\text{RF}$ chains) when the number of users is small. 
This is a limitation of the proposed scheme. However, in practice, \editbl{network operators typically set $N_\text{RF} \approx K$ to maximize the throughput of the overall system. This is \editll{exactly the scenario where} the proposed approach has significant advantage over the channel recovery based counterparts.}}
\section{Conclusion}
\label{sec:conclusion}
This paper addresses the design of hybrid analog and digital precoding matrices in a mmWave TDD massive MIMO employing single-carrier or multicarrier transmission techniques. 
The proposed learning based precoding strategy  
overcomes the limitations of the existing schemes by constructing the analog precoding matrices directly from the received pilots without the intermediate step of estimating the high dimensional channel. Further, the design of digital precoder follows from estimating a low-dimensional \editbbb{equivalent} channel using a relatively small number of pilots. In contrast to the fully direct approach \editf{of} jointly \editf{learning the analog and digital} precoding matrices, \editf{the proposed} approach \editf{significantly simplifies the} training \editf{process} and can \editf{generalize to systems with} an arbitrary number of users. 
Numerical evaluations indicate significant gains in spectral efficiency relative to the channel recovery based schemes and further demonstrate the ability of the proposed scheme to generalize in various system parameters. 

\bibliographystyle{IEEEtran}
\bibliography{IEEEabrv,referenceF}

\begin{IEEEbiography}[{\includegraphics[width=1in,height=1.25in,clip,keepaspectratio]{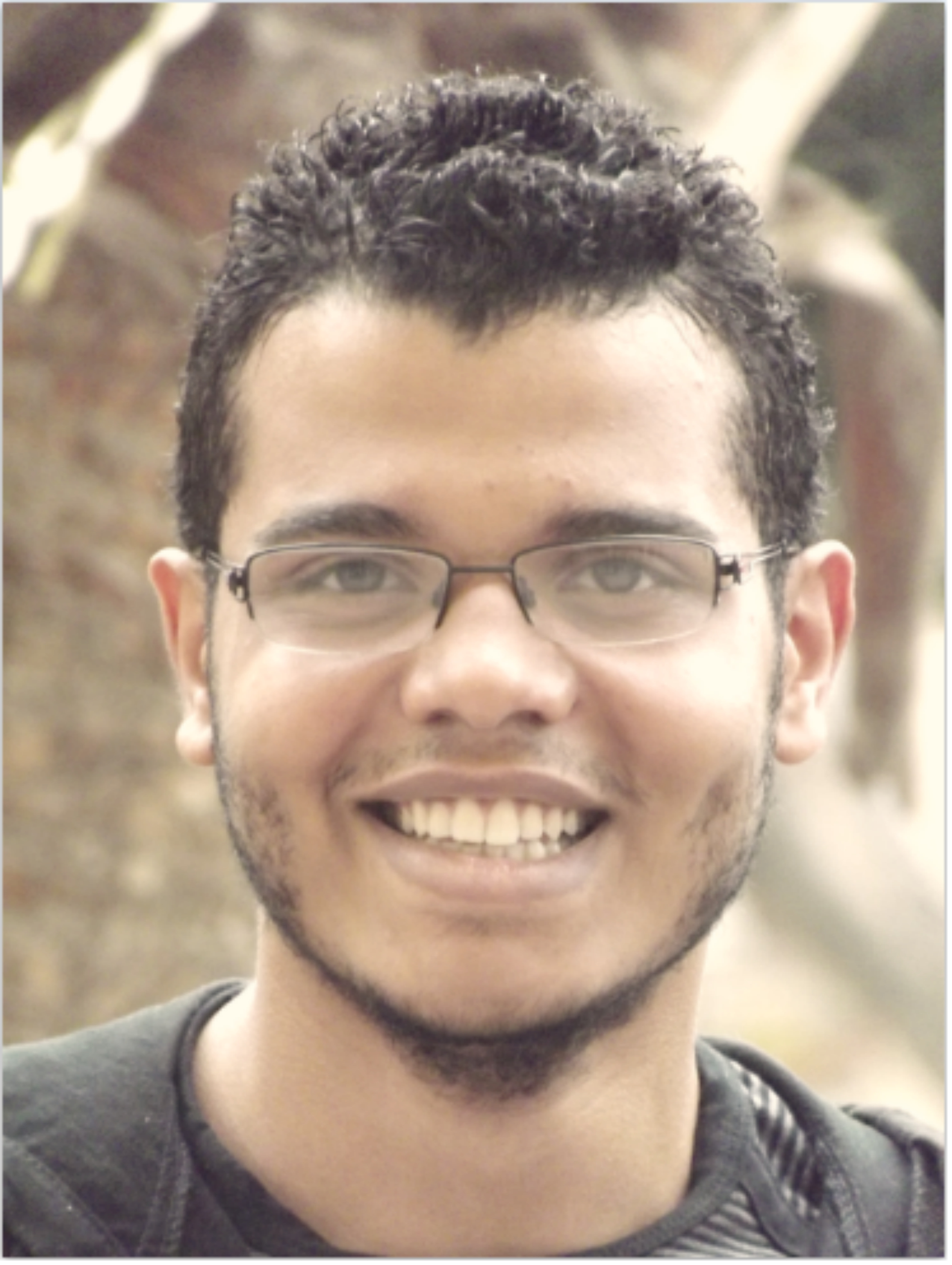}}]{Kareem M. Attiah}(Graduate Student Member, IEEE) received the B.Sc. and M.Sc. degrees in electrical engineering from Alexandria University, Alexandria, Egypt, in 2014 and 2018, respectively and is currently pursuing the Ph.D. degree with the Electrical and Computer Engineering Department, University of Toronto, Toronto, Canada, under the supervision of Prof. W. Yu. In Summer 2014, he was a Visiting Student with the Summer@EPFL Internship, École Polytechnique Fédérale de Lausanne (EPFL), Lausanne, Switzerland. In 2017, he was a member of the Carleton/Ericsson Canada Inc., a collaborative Research Project, and a Visiting Scholar with the Department of System and Computer Engineering, Carleton University. In addition, he is also a former Research Assistant with the Communications and Electronics Department, The American University in Cairo (AUC), Cairo, Egypt. His current research interests include information theory, millimeter-wave communications, machine learning, and optimization. 
\end{IEEEbiography}

\begin{IEEEbiography}[{\includegraphics[width=1in,height=1.25in,clip,keepaspectratio]{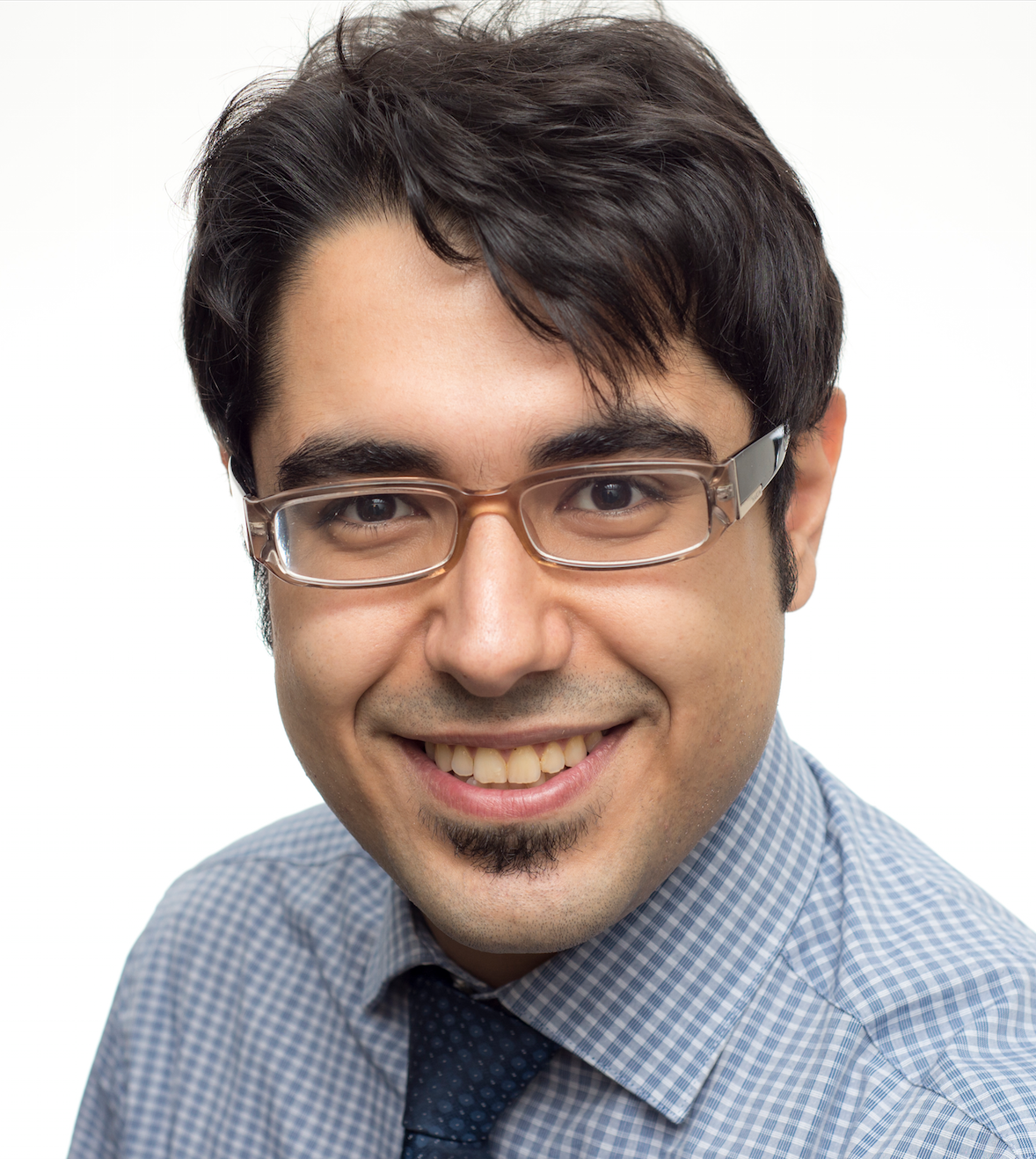}}]{Foad Sohrabi}

(Member, IEEE) received the B.A.Sc. degree from the University of Tehran, Tehran, Iran, in 2011, the M.A.Sc. degree from McMaster University, Hamilton, ON, Canada, in 2013, and the Ph.D. degree from the University of Toronto, Toronto, ON, Canada, in 2018, all in electrical and computer engineering. From March 2018 to November 2021, he was a Postdoctoral Fellow with the University of Toronto. Since October 2021, he has been working as a Senior Technical Staff at Ofinno, Reston, Virginia, USA. In 2015, he was a Research Intern with Bell Labs, Alcatel-Lucent, Stuttgart, Germany. His research interests include MIMO communications, optimization theory, wireless communications, signal processing, and machine learning. He was a recipient of the IEEE Signal Processing Society Best Paper Award in 2017.
\end{IEEEbiography}

\begin{IEEEbiography}[{\includegraphics[width=1in,height=1.25in,clip,keepaspectratio]{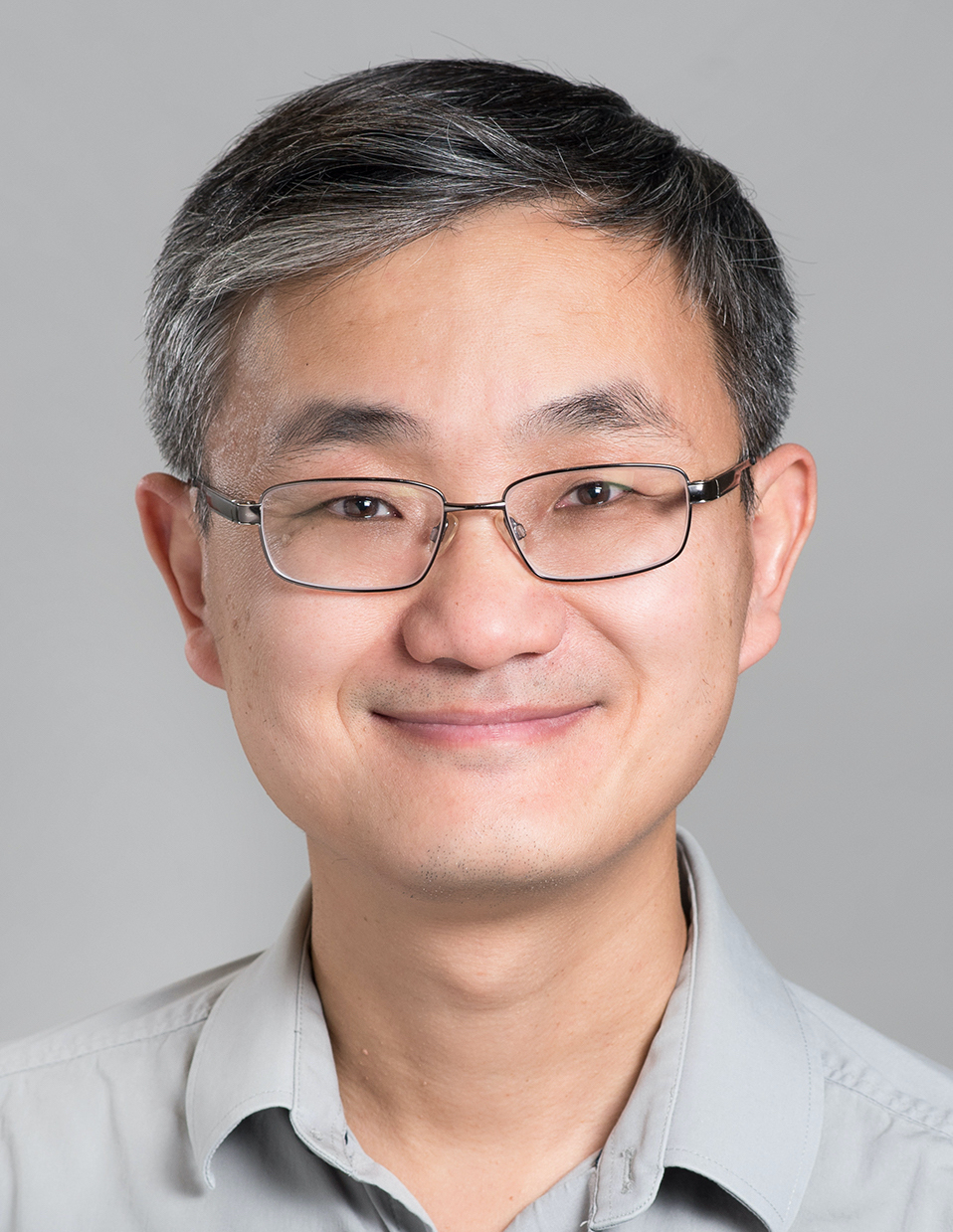}}]{Wei Yu} (Fellow, IEEE) received the B.A.Sc. degree in computer engineering and mathematics from the University of Waterloo, Waterloo, ON, Canada, in 1997, and the M.S. and Ph.D. degrees in electrical engineering from Stanford University, Stanford, CA, USA, in 1998 and 2002, respectively. Since 2002, he has been with the Electrical and Computer Engineering Department, University of Toronto, Toronto, ON, Canada, where he is currently a Professor and holds the Canada Research Chair (Tier 1) in Information Theory and Wireless Communications. He is a Fellow of the Canadian Academy of Engineering and a member of the College of New Scholars, Artists, and Scientists of the Royal Society of Canada. Prof. Wei Yu was the President of the IEEE Information Theory Society in 2021, and has served on its Board of Governors since 2015. He served as the Chair of the Signal Processing for Communications and Networking Technical Committee of the IEEE Signal Processing Society from 2017 to 2018. He was an IEEE Communications Society Distinguished Lecturer from 2015 to 2016. Prof. Wei Yu received the Steacie Memorial Fellowship in 2015, the IEEE Marconi Prize Paper Award in Wireless Communications in 2019, the IEEE Communications Society Award for Advances in Communication in 2019, the IEEE Signal Processing Society Best Paper Award in 2008, 2017 and 2021, the Journal of Communications and Networks Best Paper Award in 2017, and the IEEE Communications Society Best Tutorial Paper Award in 2015. He is currently an Area Editor of the IEEE Transactions on Wireless Communications, and in the past served as an Associate Editor for IEEE Transactions on Information Theory, IEEE Transactions on Communications, and IEEE Transactions on Wireless Communications.
\end{IEEEbiography}

\end{document}